\documentstyle[12pt,fleqn,epsfig]{article}
\begin{document}
\begin{center}
{\bf INSTITUT~F\"{U}R~KERNPHYSIK,~UNIVERSIT\"{A}T~FRANKFURT}\\
D - 60486 Frankfurt, August--Euler--Strasse 6, Germany
\end{center}

%\hfill DRAFT 4.0
\hfill IKF--HENPG/2--98

\vspace{1cm}

\begin{center}
{\Large \bf On the Early Stage }
\end{center}

\begin{center}
{\Large \bf of Nucleus--Nucleus Collisions }
\end{center}

\vspace{1cm}
\begin{center}

Marek Ga\'zdzicki\footnote{E--mail: marek@ikf.physik.uni--frankfurt.de}\\
Institut f\"ur Kernphysik, Universit\"at Frankfurt, 
Germany\\[0.8cm]

Mark I. Gorenstein\footnote{Permanent address: Bogolyubov Institute for Theoretical Physics, 
Kiev, Ukraine}$^,$\footnote{E--mail: goren@th.physik.uni-frankfurt.de}\\
Institute for Theoretical Physics, University of Frankfurt,
Germany\\
and\\
School of Physics and Astronomy, Tel Aviv University,
Israel
 
\vspace{1.5cm}

\begin{minipage}{14cm}
\baselineskip=12pt
\parindent=0.5cm
{\small 
A statistical model of the early stage of central nucleus--nucleus (A+A)
collisions is developed.
We suggest a description of the
confined state with several free 
parameters fitted to a compilation of 
A+A data at the AGS. 
For the deconfined state a simple Bag model equation
of state is assumed.  
The model  leads to the conclusion that a Quark Gluon Plasma is created
in central nucleus--nucleus collisions at the SPS. 
This result
is in  quantitative agreement with existing SPS data
on pion and strangeness production and gives a natural explanation
for their scaling behaviour.
The localization and the properties of the transition region are
discussed.
It is shown that the deconfinement transition can be detected by
observation of the
characteristic energy dependence of pion and strangeness multiplicities,
and by an increase of the  event--by--event 
fluctuations.
An attempt to understand the data on $J/\psi$ production in Pb+Pb collisions
at the SPS  within the same approach is presented.

} 

\end{minipage}

\end{center}

\vfill
\today
\newpage

\section{Introduction}

At the final state of high energy nuclear collisions many new particles
appear.
They are massive and extended objects: hadrons and hadronic resonances.
What is the nature of particle creation in strong interactions?
What is form of  matter  in 
a state of very high energy density which is created during the collision
of two nuclei?
These questions motivate a broad experimental programme in which 
nuclear collisions at high energy are investigated
\cite{QM97}.

Due to  lack of a calculable theory of strong interactions 
the interpretation of  experimental results has to rely on phenomenological
approaches.
The first models attempting to describe  high energy collisions  were 
statistical models of the early stage \cite{Fe:50, La:53},
the stage in which excitation of the
incoming matter takes place.
In their original formulations they  failed to reproduce
experimental results.
However, when a broad set of data became available
\cite{Ga:95, Ga:96}, it was realized 
\cite{Ga:95a, Ga:97} that after necessary 
generalization a statistical approach to the early stage
gives surprising agreement with the  results.
It could therefore be used as a tool to identify the 
properties of the state created at the early stage.
The aim of this paper is to further develop the statistical
model of the early stage and to apply it  to study the properties 
of the high energy density state created in nucleus--nucleus 
collisions.

A special role in this study
is played by  entropy \cite{Va:82}
(in collisions at high energy carried mainly by 
final state pions)  and heavy flavours (strangeness, charm)
 production \cite{Ko:86, Ka:86, Ma:86}.
It can be argued that they  are insensitive to the late stages
of the collision  and therefore 
carry information about  the early stage.

%The form of the strongly interacting matter at very high energy density
%is theoretically well established in terms
%of  perturbative QCD, which is in agreement with lattice QCD simulations
%at zero baryonic density.
%At high temperatures lattice QCD predicts
% a quasi--ideal gas of deconfined quarks and gluons called Quark
%Gluon Plasma (QGP) \cite{Ka:95}.
It is experimentally well established that hadrons consist of
more elementary subhadronic objects: quarks and gluons \cite{Pe:82}.
It is therefore natural to assume that at very high energy density,
higher than a typical energy density inside a hadron,
matter is in the form of a gas of subhadronic degrees of freedom called 
Quark Gluon Plasma (QGP) \cite{Co:75}.
When a quasi--ideal gas of quarks and gluons is assumed in the 
statistical model of the early stage, the results on strangeness
and pion production in nucleus--nucleus collisions at the SPS are reproduced
in an essentially parameter free way~\cite{Ga:97}.

This surprising agreement should be contrasted with the problems
\cite{Ga:95b, Je:97}
of microscopic non--equilibrium models in 
describing the same set of data.
Scaling properties of the data, natural in the thermodynamical
approach, arise in non--equilibrium models as an accidental cancelation
of many non--scaling dependences.

All that motivates further development of the experimental 
programme \cite{Ba:98}.
The main goal is to localize the collision energy region  
in which the deconfinement transition takes place and study the
properties 
of the transition itself.
In this region many anomalies are expected.
Their experimental observation will lead us to definite  conclusion
concerning the early stage of nucleus--nucleus collisions.
The corresponding experimental study at the  SPS should start in
1999 when beams of lower than maximal SPS energy will
become available.
The experiment NA49 \cite{Ba:98} is now being prepared for
this investigation. 

The main goal of this paper is 
to further  develop of the statistical model of the early
stage in order to provide  a  description 
of the transition region, with special attention paid 
to the observables which can be measured in the experimental
programme.
This can be done only when the partition function
of the confined  state is  given.
We argue that this state should not be modeled as a gas of hadrons and
hadronic resonances. 
Consequently we have to introduce an effective parametrization of the
confined state  and try to extract its properties
from the
comparison with the experimantal data.

Finally we attempt to include in the analysis the production of charm
and understand experimental data
on $J/\psi$ production \cite{Ge:98}.
The standard approach is based on the assumption  that  
$c\overline{c}$ states are created in hard QCD processes
and  later partially destroyed by interactions with
the surrounding
matter. 
This approach leads to the conclusion \cite{Sa:97}
of creation of a QGP only in central Pb+Pb collisions
at SPS, but not in the collisions of lighter nuclei.
Pion and strangeness data show however no essential
difference between central S+S and Pb+Pb collisions at the SPS.
They are consistent, within the statistical model analysis, with
the creation of a QGP already in central S+S collisions at the SPS.
Thus the crucial question is to what extent is this contradiction caused 
by the use of different approaches for data interpretation?
Can a consistent description of the data including the production of charm
be obtained using the statistical approach? 

The paper is organized as follows.
The model used further on for data interpretation and analysis of the
transition region is defined in Section 2.
The basic features of the model are presented in Section 3, 
where the approximate analytical formulae are given together with the
numerical results obtained using the full version of the model.
The model is confronted with the experimental data in Section 4.
In Section 5 the discussion of the different approaches to the
strangeness and $J/\psi$ production is given.
Summary and conclusions close the paper.

For simplicity reasons we consider only central collisions of two
identical nuclei (central A+A collisions). 
The nuclear mass number is denoted by $A$.
The whole discussion is done in the center of mass system.
The c.m. energy of  nucleon--nucleon pair and nucleon mass 
are denoted
by $\sqrt{s}_{NN}$ and $m_N$, respectively.

\newpage

\section{A Model of the Early Stage of A+A Collisions}

\vspace{0.2cm}
\noindent
1. The basic assumption of our model is that the 
production of new degrees of freedom 
in the early stage of nucleus--nucleus collisions is a statistical process.
Thus formation of all microscopic
states allowed by conservation laws
is equally probable.
This means that the probability to produce a given macroscopic
state is proportional to the total number of its microscopic
realizations, i.e.
a macroscopic state 
probability $P$ is
\begin{equation}
P \sim e^S,
\end{equation}
where $S$ is the entropy of the macroscopic state.

\vspace{0.2cm}
\noindent
2. As the particle creation process does not produce net baryonic,
flavour and
electric charges only states  with the total
baryon, flavour and electric
numbers equal to zero should be considered.
Thus the properties of the created state are entirely defined by the
volume in which production takes place, 
the  available energy  and a
partition function.  
In the case of collisions of large nuclei
the  thermodynamical approximation can be used and
 the dependence on the volume and the energy
reduces to the dependence on the energy density.
The state properties can be given in the form of an equation of state.

\vspace{0.2cm}
\noindent
3. We assume that the creation of the
early stage entropy in central A+A collisions  takes
place in the volume equal to  the Lorentz contracted
volume occupied by the colliding nucleons (participant nucleons) 
from a single nucleus:
\begin{equation}\label{volume}
V = \frac {V_0} {\gamma},
\end{equation}
where $V_0 = 4/3 \pi r_0^3 A_p$ and $\gamma = \sqrt{s}_{NN}/(2 m_N)$
and $A_p$ is the number of participant nucleons from a single nucleus.
The $r_0$ parameter is taken to be 1.30 fm in order to fit the mean
baryon density in the nucleus, $\rho_0 = 0.11$ fm$^{-3}$.

\vspace{0.2cm}
\noindent
4. Only a fraction of the total 
energy in A+A collision  is transformed
into the energy of 
new degrees of freedom
created in the early stage.
This is because
a part of the energy is carried by the  net baryon number which
is conserved during the
collision. 
The released energy can be expressed as:
\begin{equation}\label{energyin}
E = \eta (\sqrt{s}_{NN} - m_N)~A_p~.
\end{equation}
The parameter $\eta$ 
defines the  fraction of the available energy
used in the production process. It  
is assumed to be independent of
the collision energy and the system size for A+A collisions 
discussed in this paper.
This assumption is in agreemnet with the experimental data
providing that a correction for pion absorption effects 
(see point 15 below) is taken into
account.
It   is usually justified by 
quark--gluon structure
of the nucleon \cite{Po:74}. 
The value of  $\eta$ used for the numerical calculations
is 0.67 (see Section 4 for details).

\vspace{0.2cm}
\noindent
5. In order to predict a probability of creation of a given 
macroscopic state all possible degrees of freedom
and interaction between them should be given
in the form of the partition function.
In the case of large enough volume the grand canonical approximation 
can be used and the state properties can be given in the form of
an equation of state.  
The question of how one can use this equation of state 
to calculate the space--time evolution (hydrodynamics)
of the created system requires a  separate study. 
 
\vspace{0.2cm}
\noindent
6. The most elementary particles of strong interaction are quarks
and gluons.
In the following we consider $u$, $d$ and $s$ quarks and the corresponding
antiquarks with the internal number of degrees of freedom equal to
6 (3 colour states $\times$ 2 spin states). In the entropy evaluation the
contribution of $c$, $b$ and $t$ quarks can be neglected due to 
their  large masses. 
The charm production is discussed separately in Section 5.
The internal number of degrees of freedom for gluons is 16
(8 colour states $\times$ 2 spin states).
The masses of gluons and nonstrange (anti)quarks are taken to be 0,
the strange (anti)quark mass  is taken to be 175 MeV \cite{Le:96}. 
 
\vspace{0.2cm}
\noindent
7. In the case of creation of colour quarks and gluons the
equation of state is assumed to be
the ideal gas equation of state modified by the bag constant $B$ in order
to account for 
the strong interaction between quarks and gluons
and the surrounding  vaccum (see e.g. \cite{bag}):
\begin{equation}
p = p^{id} - B~,~~~
\varepsilon = \varepsilon^{id} + B,
\end{equation}
where $p$ and
$\varepsilon$ denote pressure and energy density, respectively,
and $B$ is the so called bag constant.
The equilibrium state  
defined above is  called
Quark Gluon Plasma or
Q--state.

\vspace{0.2cm}
\noindent
8. At the final freeze--out stage of the collision 
the degrees of freedom are hadrons~-- 
extended and massive objects composed of (anti)quarks and
gluons.
Due to their finite proper volume hadrons can exist in their well defined
asymptotic states only at rather low energy densities.
Estimates $\epsilon < 0.1\div0.4$~GeV/fm$^3$ for the hadron gas with van der
Waals excluded volume have been found in Ref.~\cite{Ye:97}.
 In the early stage of A+A collisions such  low energy density is  
created only at very low collision energies of a few GeV per
nucleon.
We note also that 
asymptotic hadronic states can be questioned as possible degrees
of freedom in the early stage on the basis of our current
understanding of $e^+ + e^-$ anihilation processes, where
the initial degrees of freedom are found to be colourless
$q\overline{q}$ pairs~\cite{Pe:82}.
The above statements lead us to the conclusion that there is no
satisfactory model of the confined state at the early stage
of A+A collisions at the AGS.

Guided by these considerations we introduce an effective parametrization
of the confined state.
We assume that at collision energies lower than the energy needed
for a QGP creation
the early stage effective degrees of freedom can be approximated
by 
point--like colourless  bosons.
This state is denoted as W--state (White--state).
The nonstrange degrees of freedom which dominate 
the entropy production
are taken to be massless, as seems to be suggested by the
original analysis of the
entropy production in N+N and A+A collisions \cite{Ga:95a}.
Their internal number of degrees of freedom was fitted to the same
data
\cite{Ga:95a, Ga:97} 
to be about 3 times
lower than the
internal number of effective
degrees of
freedom in A+A collisions at SPS, where in our model
creation of QGP takes place.
The internal number of degrees of freedom for a QGP is
16 + (7/8)$\cdot$36 $\cong$ 48 and therefore the internal number of
nonstrange degrees of freedom for low energy collisions is taken
to be 48/3 = 16.
The mass of strange degrees of freedom is assumed to be 500 MeV, 
equal to the kaon mass.
The internal number of strange degrees of freedom is estimated to
be 14 as suggested by the fit to the strangeness and pion data at the AGS
(see Section 4).
The phenomenological reduction factor 3 is used 
in our numerical calculations  between the total number of
degrees of freedom for Q--state and nonstrange degrees of freedom of
 W--state
because of the different magnitude of strangeness suppression 
due to different
masses of strangeness carriers in both cases.
The ideal gas equation of state is selected.
We would like to underline once more  that the above description
of the confined  state should be treated only
as an effective parametrization. Its parameters are fixed by fitting 
A+A data at the AGS. 
This parametrization is needed for the
extrapolation to higher collision energies where
the transition between the confined and deconfined state
is expected.
It is, however, intresting to speculate (see Appendix A) about 
the possible physical meaning of 
the obtained parameters of
the degrees of freedom in the W--state.

\vspace{0.2cm}
\noindent
9.  
For large enough  volume the grand canonical
approximation can be used and the calculation of the entropy
is significantly simplified.
In a large system  only one macroscopic state is
produced -- the state with the maximum entropy density, $s$.
This is because the relative probability of the state with the
entropy density $s^{\prime} < s$ 
is given by:
\begin{equation}
\frac {P} {P_{MAX}}~ =~ \exp~[~V~(s^{\prime}~ -~s)~]~.
\end{equation}
Thus 
the relative probability decreases to zero when the
volume increases to infinity for any value of  $s^{\prime} < s$. 

\vspace{0.2cm}
\noindent
10. In the case of finite (small) volume  the conservation laws
should be accounted for in a strict way (canonical or microcanonical 
treatment).
The macroscopic states with an entropy density lower than the maximum
one
are created with final probabilities.
As the physical properties of various states can be significantly 
different (see Section 3) sizeable nontrivial event--by--event
fluctuations are expected.

\vspace{0.2cm}
\noindent
11. The maximum entropy state is called equilibrium state.
In the model with two different states (W and Q) the form of 
the maximum entropy state changes with the collision energy.
The regions, in which the equilibrium state is in the form of a pure
W or a pure Q state, are separated by the region in which both states
coexist (mixed phase or W--Q--state).

\vspace{0.2cm}
\noindent
12. It is important to note that the formation of a state  
in  global equilibrium in the early stage of nuclear 
collisions is a consequence of our basic assumption
that all possible microscopic states are created with equal
probability. 
Thus it is due to the assumed statistical nature of the primary creation
process and it is not due to equilibration by a long lasting sequence
of secondary interactions.

\vspace{0.2cm}
\noindent
13. The globally equilibrated state created in the early stage 
expands and finally freezes--out into hadrons and
hadronic resonances.
Recent analysis suggests that this hadronization process can be
described by a statistical model \cite{Be:96, Be:97}.
Thus phase--space seems to govern not only the production 
of entropy and the flavour content of the state in the early
stage, as discussed in this paper, but also
its conversion to hadrons which happens at a significantly lower
energy scale.

\vspace{0.2cm}
\noindent
14. Note that due to the Lorentz contraction 
the shape of the early stage volume is non--spherical.
This causes  the isotropic angular distribution of particles
in the early stage to be converted during an anisotropic expansion
into a forward--backward peaked distribution as
observed in the experimental data.

\vspace{0.2cm}
\noindent
15. We assume that the only process which changes the entropy content
of the produced matter during the expansion, hadronization and
freeze--out is an interaction with the baryonic subsystem.
It was argued that it leads to  an entropy transfer to baryons
which corresponds to the effective absorption of about 
0.35 $\pi$--mesons per baryon
\cite{Ga:95a, Ga:96a}.
This interaction causes also that the produced hadrons in the final
state do not obey symmetries of the early stage production
process, i.e. 
a final hadronic state has non--zero
baryonic number and electric
charge.

\vspace{0.2cm}
\noindent
16. It is assumed that the total number of $s$ and $\overline{s}$
quarks created in the early stage
is conserved during the expansion, hadronization and 
freeze--out.

\newpage

\section{Calculations}

In the first part of this section we analyze the 
simplified version of the model which allows us
to perform calculations in an analytical way.
The results of the numerical calculations done within the
full version of the model are presented in the second
part of the section. 

The  calculation are performed
in the grand canonical formulation
which is justified for large enough systems discussed in this paper.  
All chemical potentials have to be equal to zero, 
as we consider only systems with  all conserved charges equal to zero.
Thus
the temperature $T$ remains the only independent thermodynamical
variable in the thermodynamical limit when the system volume
goes to infinity. 
It is convenient to define
the system equation of state
in terms of the pressure function
$p=p(T)$ 
as the entropy and energy densities
can be caluclated from the thermodynamical
relations:
\begin{equation}\label{termid}
s(T)~=~\frac{dp}{dT}~,~~~~\varepsilon (T)~=~T\frac{dp}{dT}~-~p~.
\end{equation}
In the case of an ideal gas the pressure of the 
particle species `$j$' is given by:
\begin{equation}\label{pressi}
p^j(T)~=~\frac{g^j}{2\pi^2}\int_{0}^{\infty}k^2dk~\frac{k^2}
{3(k^2+m_j^2)^{1/2}}~\frac{1}{\exp\left(\frac{\sqrt{k^2+m_j^2}}{T}\right)
~\pm ~1}~,
\end{equation}
where $g^j$ is the internal number of degrees of freedom (degeneracy
factor) for $j$--th species, $m_j$ is a particle mass, 
`--1' appears in Eq.~(\ref{pressi}) for bosons and `+1' for fermions. 
The pressure $p(T)$ for an ideal gas of several particle
species is additive: $p(T)=\sum_j p^j(T)$. The same is
valid for the entropy and energy densities (\ref{termid}).

\subsection{Analytical Calculations}

In order to perform analytical calculations  
of the system entropy and illustrate the model
properties  we simplify our consideration
assuming that all degrees of freedom are massless.
In this case the pressure function 
(Eq.~(\ref{pressi})) is equal to:
\begin{equation}\label{presso}
p^j(T)~=~\frac{\sigma^j}{3}~T^4~,
\end{equation}
where $\sigma^j$ is
the so called Stephan--Boltzmann constant, equal to $\pi^2g^j/30$
for bosons and $\frac{7}{8}\pi^2 g^j/30$
for fermions. 
The total pressure in the ideal gas of several 
massless species can be presented then as $p(T)=\pi^2 g T^4/90$
with the effective
number of degrees of freedom $g$ 
given  by 
\begin{equation}\label{g}
g~=~g^{b}+\frac{7}{8}~g^{f},
\end{equation}
where $g^b$ and $g^f$ are internal degrees of freedom
of all bosons and fermions, respectively.
The $g$ parameter  is taken to be
$g_W$ for W--state and
$g_Q$ for Q--state, with $g_Q >  g_W$.

The pressure, energy and entropy densities are given then as:
\begin{equation}\label{wmatter}                    
p_W(T)=\frac{\pi^2g_W}{90}~T^4~,~~~
\varepsilon_W(T)=\frac{\pi^2g_W}{30}~T^4~,~~~
s_W(T)=\frac{2\pi^2g_W}{45}~T^3~,\\ 
\end{equation}
\begin{equation}\label{qmatter}
p_Q(T)=\frac{\pi^2g_Q}{90}~T^4-B~,~~~
\varepsilon_Q(T)=\frac{\pi^2g_Q}{30}~T^4+B~~~,~
s_Q(T)=\frac{2\pi^2g_Q}{45}~T^3~, 
\end{equation}
for the pure W-- and Q--state, respectively.
Note the presence of the non--perturbative bag terms 
in addition to the ideal quark--gluon gas expressions for
the pressure and energy density of the Q--state.

The 1st order phase transition 
between W-- and Q--state is defined by the Gibbs criterion
\begin{equation}\label{ptr}
p_W(T_c)~=~p_Q(T_c)~,
\end{equation}
from which  the phase transition temperature can be calculated as:
\begin{equation}\label{tcr}
T_c~=~\left[\frac{90B}{\pi^2(g_Q-g_W)}\right]^{1/4}~.
\end{equation}
At $T=T_c$ the system is in the {\it mixed} phase with
the energy and entropy densities given by 
\begin{equation}\label{mixed}
\varepsilon_{mix}=(1-\xi)\varepsilon_W^c~+~\xi 
\varepsilon_Q^c~,~~~~
s_{mix}=(1-\xi)s^c_W~+~\xi s_Q^c~,
\end{equation}
where $(1-\xi)$ and $\xi$ are the relative volumes 
occupied by the W-- and Q--state, respectively.
~From Eqs.~(\ref{wmatter}, \ref{qmatter}) one finds
the energy density discontinuity (`latent heat')
\begin{equation}\label{lheat}
\Delta \varepsilon ~\equiv ~ \varepsilon_Q(T_c)-
\varepsilon_W(T_c)~\equiv ~\varepsilon_Q^c-
\varepsilon_W^c~=~4B~.
\end{equation}

In our model 
the early
stage energy density is an increasing function of
the collision energy and it is given by  
(see Eqs. ( \ref{volume}, \ref{energyin})): 
\begin{equation}\label{endensity} 
\varepsilon ~ \equiv~ \frac {E}
{V} ~=~\frac{\eta~\rho_0~(\sqrt{s}_{NN}-2m_N)~\sqrt{s}_{NN}}{2 m_N}~. 
\end{equation} 
According to our basic assumption (point 1 in Section~2)
the created macroscopic state should be defined 
by the entropy density maximum condition:
\begin{equation}\label{extremum}
s(\varepsilon)~=~\max~\{~s_W(\varepsilon),~
s_Q(\varepsilon),~s_{mix}(\varepsilon)~\}~.
\end{equation} 
In Appendix B we prove a remarkable equivalence of the Gibbs criterion
(largest pressure function $p_i$ in the pure $i$--phase
and equal pressures (\ref{ptr}) in the mixed phase) and
the maximum entropy criteria (\ref{extremum})
for an arbitrary equation of state $p=p(T)$ 
with a 1-st order phase transition. 
From this fundamental equivalence it follows that
for $\varepsilon < \varepsilon_W^c$ or
$\varepsilon > \varepsilon_Q^c$ the system consists of pure W-- or
Q--state, respectively, with entropy density given by the following
equations:
\begin{equation}\label{entropyw}
s_W(\varepsilon)~=~\frac{4}{3} \left(\frac{\pi^2 g_W}{30}\right)^{1/4}~
\varepsilon^{3/4}~,
\end{equation}
\begin{equation}\label{entropyq}
s_Q(\varepsilon)~=~\frac{4}{3}\left(\frac{\pi^2 g_Q}{30}\right)^{1/4}~
(\varepsilon -B)^{3/4}~.
\end{equation}
For $\varepsilon_W^c<\varepsilon<\varepsilon_Q^c$
the system is in the mixed phase~(\ref{mixed}) and its  entropy
density can be expressed as:
\begin{equation}\label{entropym}
s_{mix}(\varepsilon)~=~ \frac{\varepsilon_Q^cs_W^c - \varepsilon_W^cs_Q^c}
{4B}~+~\frac{s_Q^c - s_W^c}{4B}~\varepsilon~
\equiv~a~+~b~\varepsilon ~.
\end{equation}
The ratio of the total entropy 
of the created state to the number of nucleons paticipating in A+A
collisions is given as
\begin{equation}\label{Entropy}
\frac{S}{2A_p}~=~ \frac{V~s}{2A_p}~=~\frac{
m_N~s}{\rho_0\sqrt{s}_{NN}}
\end{equation}
and it is independent on the number of participant nucleons.
The entropy density $s$ in Eq.~(\ref{Entropy}) is given
by our general expressions (\ref{extremum}) with
$\varepsilon$ defined by Eq.~(\ref{endensity}). 
For small $\sqrt{s}_{NN}$ the 
energy density (\ref{endensity}) corresponds to
the pure W--state and one finds
\begin{equation}\label{Entropyw} 
\left(\frac{S}{2A_p}\right)_{W}~=~C~g_W^{1/4}~F~,
\end{equation}
where
\begin{equation}\label{c}
C~=~\frac{2}{3}\left
(\frac{\pi^2m_N}{15\rho_0}\right)^{1/4}~\eta^{3/4}~,~~~~
F~ =~ \frac { (\sqrt{s}_{NN} - 2 m_N)^{3/4} }  { (\sqrt{s}_{NN})^{1/4} }~.
\end{equation}
Thus for low collision energies, where the W--state is created, the 
entropy per participant nucleon is proportional to $F$.
For high $\sqrt{s}_{NN}$ the pure Q--state is formed
and Eq.~(\ref{Entropy}) leads to
\begin{eqnarray}\label{Entropyq}
\left(\frac{S}{2A_p}\right)_Q~&=&~C~g_Q^{1/4}~F~\left(1~-~
\frac{2m_N B}{\eta \rho_0 (\sqrt{s}_{NN}-2m_N)\sqrt{s}_{NN}}
\right)^{3/4} \\
\nonumber
& \cong&~C~g_Q^{1/4}~F~\left(1-\frac{3m_NB}{2\eta\rho_0
F^4}\right)~.
\end{eqnarray}
For large values of $F$ the entropy per participant nucleon in Q--state
is also proportional to $F$.
The slope 
is, however, larger than the corresponding slope for  the W--state
by a factor $(g_Q/g_W)^{1/4}$.
In the interval of $F$ in which the mixed phase is formed the energy 
dependence of the entropy per participant nucleon is given by:
\begin{equation}\label{Entropym}
\left(\frac{S}{2A_p}\right)_{mix}~=~ \frac{C_1}{\sqrt{s}_{NN}}~+~
C_2~(\sqrt{s}_{NN}-m_N)~,
\end{equation}
where
\begin{equation}\label{c1}
C_1~=~\frac{m_N}{\rho_0}~a~,~~~~ C_2~=~\eta ~b~.
\end{equation}
Eq.~(\ref{Entropym}) gives approximately a $F^2$ increase
of the entropy per participant nucleon in the mixed phase region.

Let us now turn to strangeness and assume that 
$g^s_W$ and $g^s_Q$ are 
the numbers of internal degrees of freedom of (anti)strangness 
carriers in W-- and Q--state, respectively.
The total entropy of the considered state is given by a sum of entropies
of strange and nonstrange degrees of freedom.
Provided that all particles are massless the fraction of entropy
carried by strange (and antistrange) particles is proportional
to the number of strangeness degrees of freedom:
\begin{equation}\label{strentr}
S_s ~=~ \frac {g^s} {g}~ S~.
\end{equation}
Eq.~(\ref{strentr}) is valid for both W-- and Q--state.  
Note that all degeneracy factors are calculated according
to the general relation (\ref{g}).
For massless particles of the $j$--th
species the entropy is proportional to the
particle number
\begin{equation}
S_j ~= ~4 N_j~.
\end{equation}
Thus the number of strange and antistrange particles can
be expressed as
\begin{equation}
N_s + N_{\overline{s}} ~=~ \frac {S}{4}~ \frac {g^s} {g}~,
\end{equation}
and the strangeness to entropy ratio is equal to
\begin{equation}\label{strent}
\frac { N_s + N_{\overline{s}} } {S}
~ =~ \frac {1}{4}~ \frac  {g^s} {g}~.
\end{equation}
We conclude therefore that
the strangeness to entropy ratio for the ideal gas of massless
particles is dependent only on the ratio of strange  to all 
degrees of freedom, $g^s/g$. 
This ratio is expected to be equal to
$g_Q^s/g_Q\cong 0.22$ in Q--state and $g_W^s/g_W\cong 0.5$
in W--state (see the next subsection). 
Therefore a phase transition from W--
to Q--state
should lead to a decrease of the strangeness to entropy
ratio by a factor of about 2. This simple picture will be modified
essentially because of the large 
value of the mass of strange
degrees of freedom in W--state ($m_W^s\cong500$ MeV)
in comparison to $T$.
In this case
the left hand side of Eq.(\ref{strent}) is
a strongly increasing  function of $T$.
The right hand side of Eq.(\ref{strent})  gives
then only its asymptotic value approached for  $T>>m_W^s$. 
The numerical calculations
for the selected parameters of W-- and Q--state are given below.

\subsection{Numerical Calculations}

The results of the calculations performed within the full
version of the model as defined in  Section 2 are presented below.
As all nonstrange degrees of freedom are assumed to be massless,
their thermodynamical functions obtained in
the previous subsections 
can be used.
For the number of nonstrange degrees of freedom we get: 
\begin{equation}\label{gnonstrange} 
g_Q^{ns}~=~2\cdot8~+~\frac{7}{8}\cdot 2\cdot 2\cdot 3\cdot 2 ~=~37~;
~~~~~~g_W^{ns}~=~16~.
\end{equation}
The strange degrees of freedom are considered as massive ones.
The Eq. (\ref{pressi}) is used with
\begin{equation}\label{gstrange}
g_Q^s~=~2\cdot 2\cdot 3  ~=~12~,~~m_Q^s~\cong ~ 175~ \mbox{MeV}~;
~~~~~~ g_W^s~=~14~,~~m_W^s~\cong~ 500~ \mbox{MeV}~.
\end{equation}
Note that there is no factor `7/8' in the expression for
$g_Q^s$ (\ref{gstrange})
as the Eq. (\ref{pressi}) with Fermi momentum distribution
is taken. 
The contributions of strange degrees of freedom to
the entropy and energy densities are calculated using
thermodynamical relations (\ref{termid}).

\begin{figure}[p]
\epsfig{file=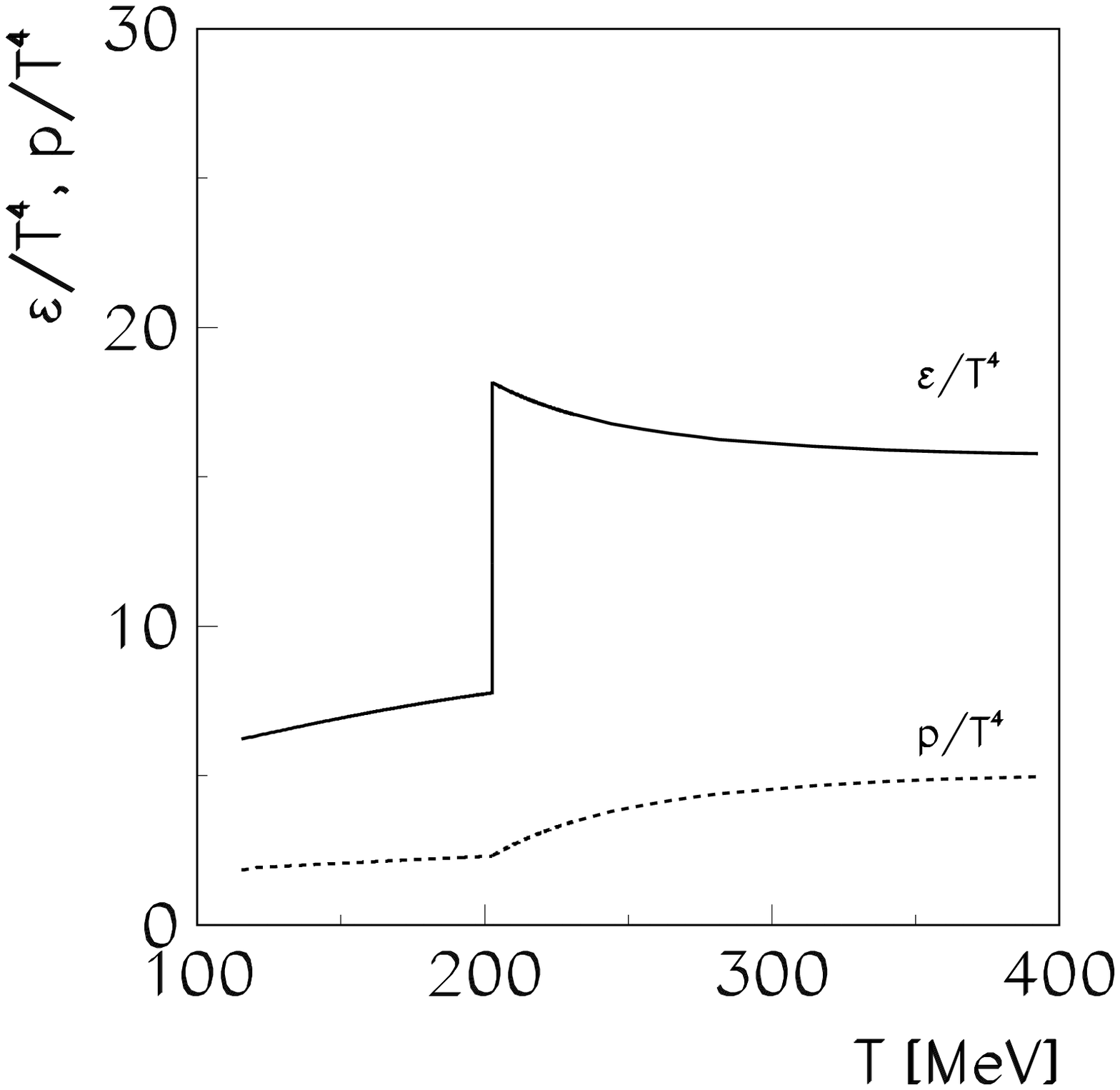,width=14cm}
\caption{
Energy density and pressure divided by $T^4$ as a function of temperature
$T$. 
}
\label{ept4}
\end{figure}

 In order to demonstrate properties of the equation of state
the ratios of $\varepsilon/T^4$ and  $s/T^4$ 
are plotted in Fig. \ref{ept4} as a function
of the temperature.
The bag constant $B$ = 600 MeV/fm$^3$ was adjusted  such 
that the critical temperature, $T_c$ is equal to
200 MeV.
This choice of $T_c$  
was suggested by the results of the analysis of hadron 
multiplicities in A+A collisions at SPS energies.
They indicate that the hadron chemical freeze--out (or hadronization)
occurs at a temperature of 160--190 MeV \cite{Sa:93, Br:95, Ri:97, Be:96,
Be:97}.  

\begin{figure}[p]
\epsfig{file=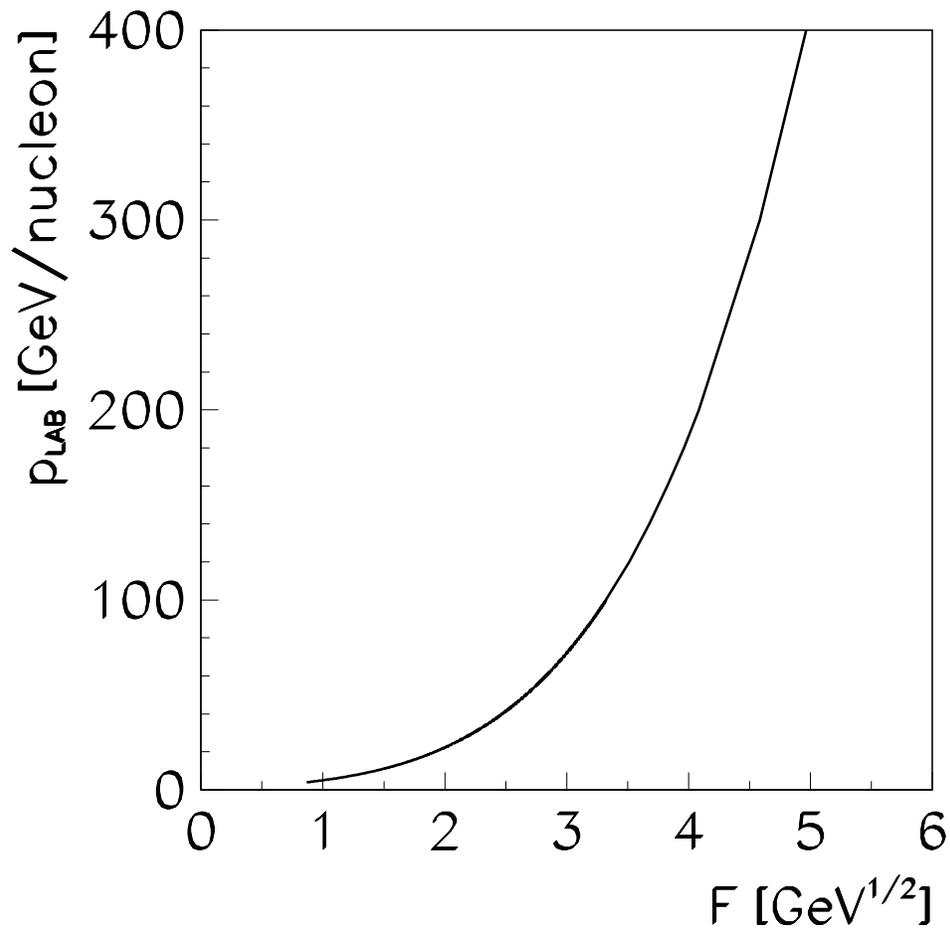,width=14cm}
\caption{
Relation between laboratory momentum per nucleon and the
Fermi--Landau energy variable $F$.
The values of $F$ for $p_{LAB}$ = 5, 10, 15, 40, 80, 160 and
200 A$\cdot$GeV are 0.99, 1.43, 1.71, 2.47, 3.10,
3.82 and 4.08 GeV$^{1/2}$, respectively.
}
\label{plab}
\end{figure}

As pointed out in the previuos section a convenient
variable to study collision energy dependence is the 
Fermi--Landau variable $F$.
This variable is  used for the further analysis.
The relation of  the $F$ variable to the laboratory
momentum $p_{LAB}$ is shown in Fig. \ref{plab}.
The values of $F$ for the top SPS and AGS energies are about
4 GeV$^{1/2}$ and 1.7  GeV$^{1/2}$, respectively.

\begin{figure}[p]
\epsfig{file=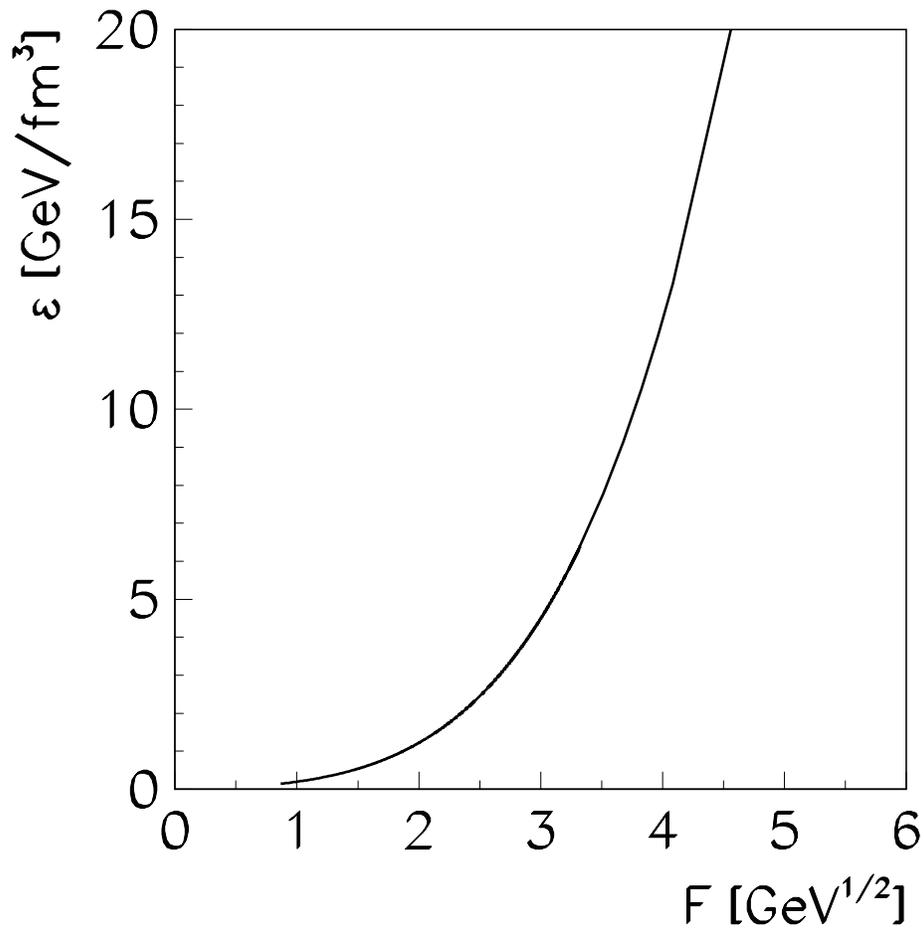,width=14cm}
\caption{
The early stage energy density as a function of $F$.
The values of $\epsilon$ for 
$F$ = 0.99, 1.43, 1.71, 2.47, 3.10,
3.82 and 4.08 GeV$^{1/2}$
are 0.20, 0.47, 0.77, 1.71, 2.36, 5.03, 10.53, and 13.32
GeV/fm$^3$, respectively.
}
\label{eden}
\end{figure}

The energy density can be calculated in a unique way
on the base of assumptions  `3' and `4' 
from  Section 2. 
The energy density (\ref{endensity}) obtained in this way is plotted
in Fig. \ref{eden} as a function of $F$.
The energy densites for the SPS and AGS energies are about
12 GeV/fm$^3$ and 0.7 GeV/fm$^3$, respectively.

\begin{figure}[p]
\epsfig{file=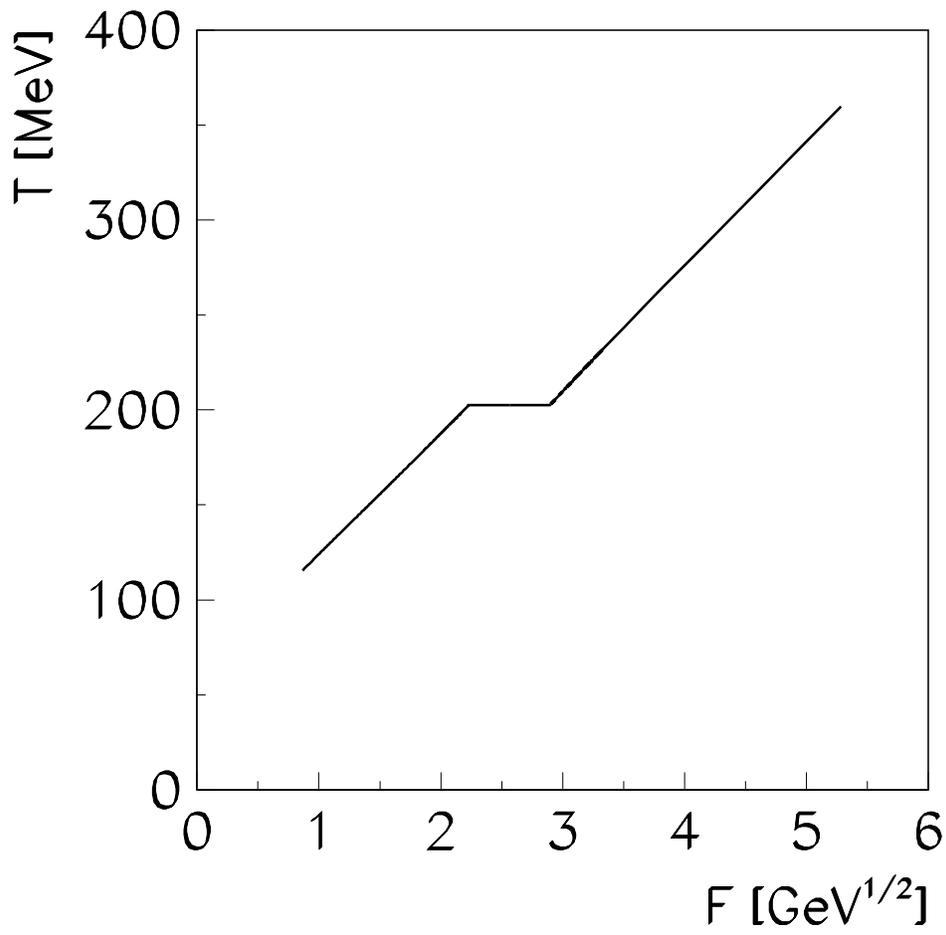,width=14cm}
\caption{
The early stage temperature as a function of $F$.
The values of $T$ for 
$F$ = 0.99, 1.43, 1.71, 2.47, 3.10,
3.82 and 4.08 GeV$^{1/2}$
are 123, 151, 169, 203, 217, 264  and 281 MeV, 
respectively.
}
\label{temp}
\end{figure}

 The dependence of the early stage temperature $T$ on
$F$ is shown in Fig. \ref{temp}.
Outside the transition region $T$ increases in an approximately
linear way.
Inside the transition region $T$ is constant ($T = T_c = 200$)
MeV.  
The transition region begins at $F= 2.23$ GeV$^{1/2}$ 
($p_{LAB}$ = 30 A$\cdot$GeV) and
ends at $F= 2.90$ GeV$^{1/2}$ ($p_{LAB}$ = 64 A$\cdot$GeV).

\begin{figure}[p]
\epsfig{file=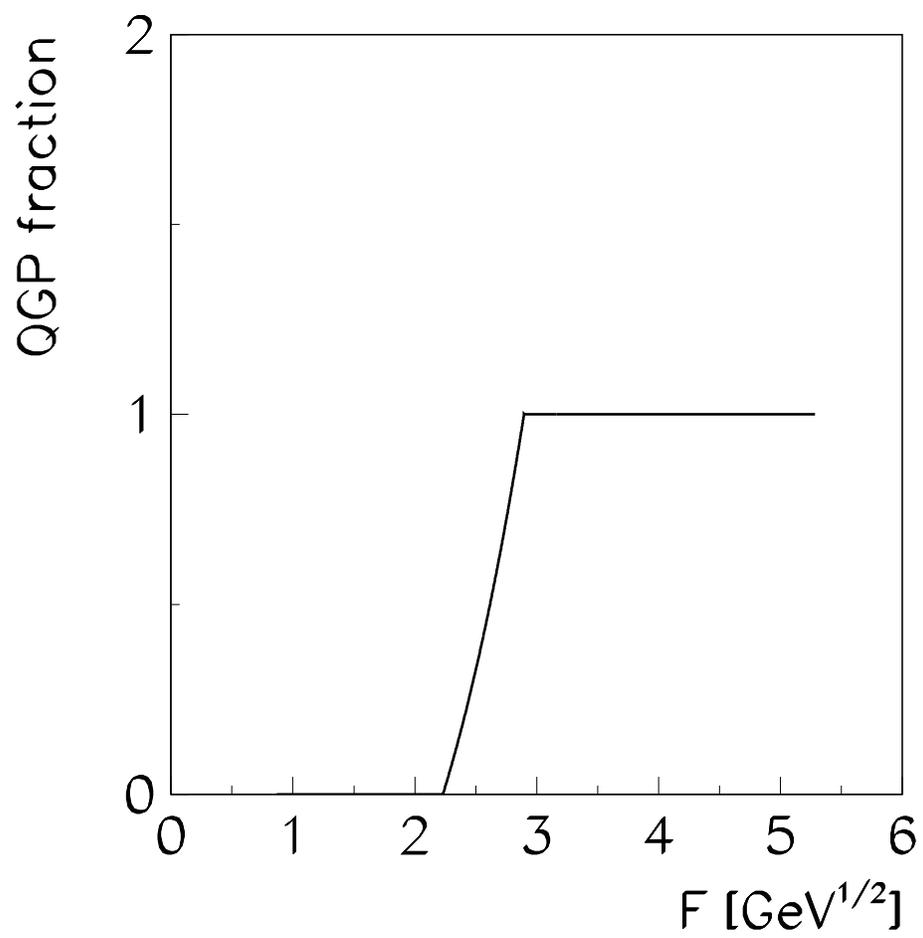,width=14cm}
\caption{
The fraction of volume occupied by a QGP as a function of $F$. 
}
\label{ksi}
\end{figure}

The fraction  of the volume occupied by the Q--state, $\xi$, increases
rapidly in the transition region, as shown in Fig. \ref{ksi}.

\begin{figure}[p]
\epsfig{file=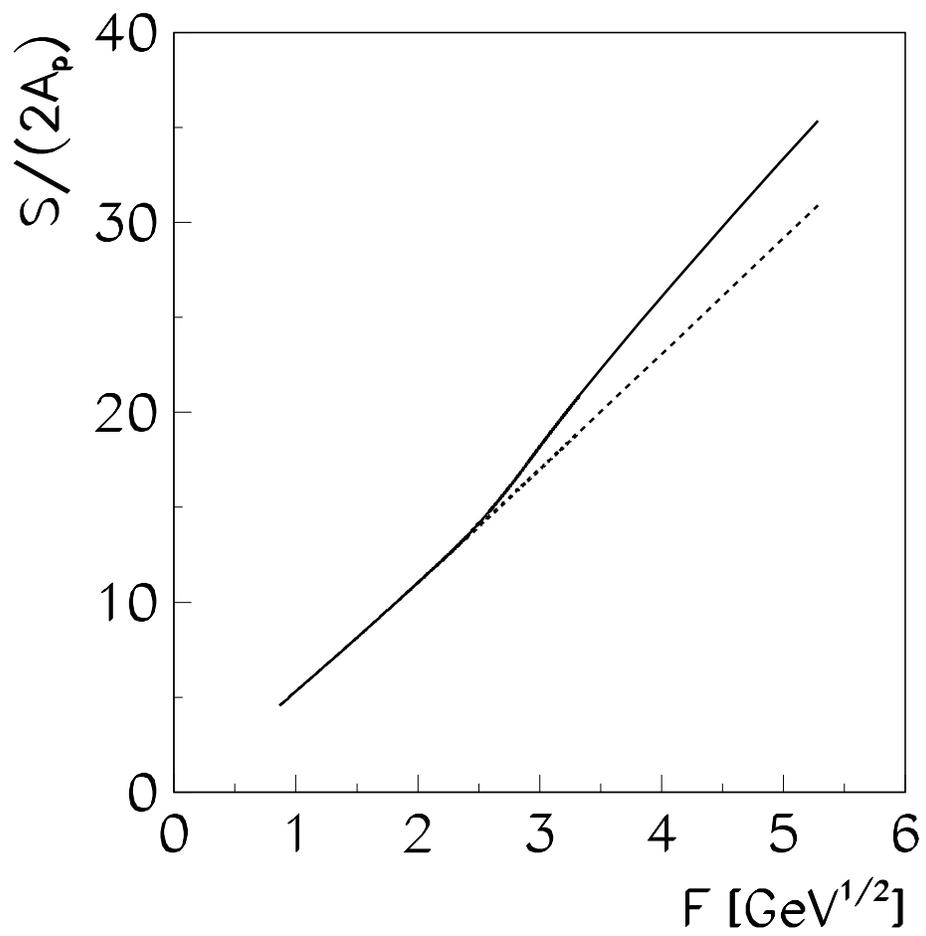,width=14cm}
\caption{
The entropy per participant nucleon as a function of $F$ (solid line).
Dashed line indicates  the dependence obtained assuming 
that there is no transition to the QGP.
}
\label{spb}
\end{figure}

\begin{figure}[p]
\epsfig{file=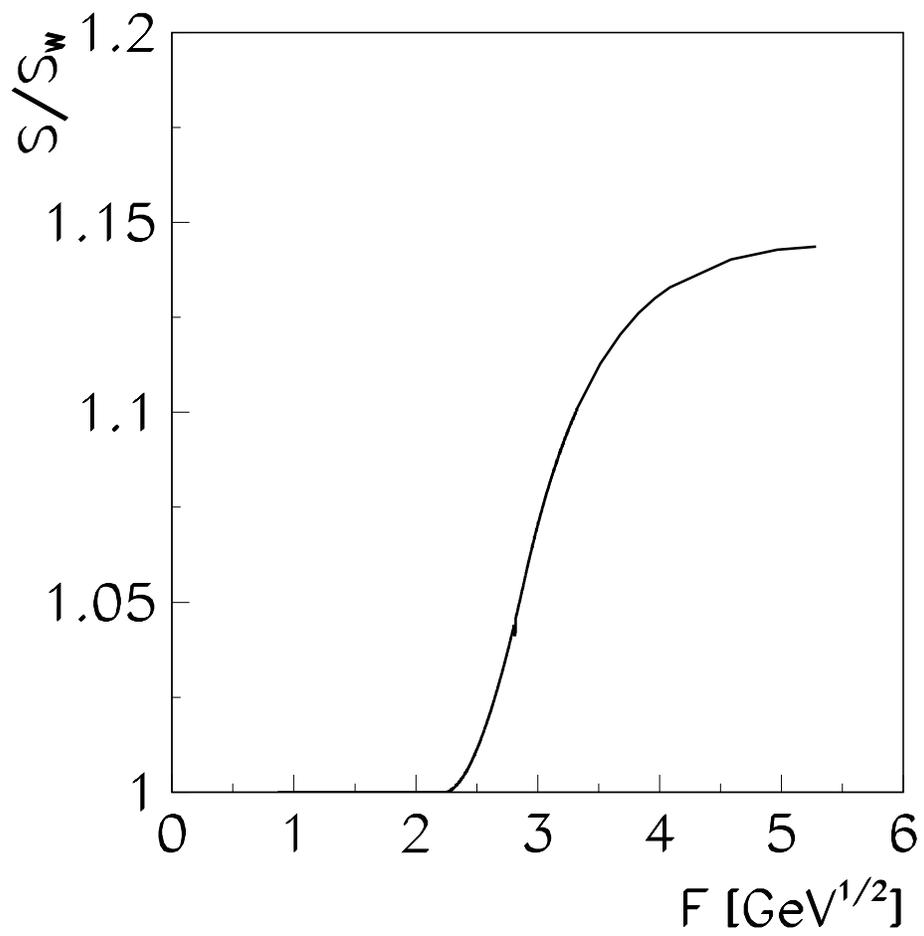,width=14cm}
\caption{
The ratio between the entropy calculated within our model
and the entropy obtined assuming absence of the phase transition 
to the QGP.
}
\label{ratio}
\end{figure}

The dependence of the entropy per participant nucleon  on 
$F$ is shown in Fig. \ref{spb}.
Outside the transition region the entropy increases  approximately
proportionaly to $F$, but the slope in the Q--state region is larger than
the slope in the  W--state region.
The ratio between the value of entropy obtained in our model
and the entropy calculated assuming that only  W--state exists is shown
in Fig.~\ref{ratio}.

\begin{figure}[p]
\epsfig{file=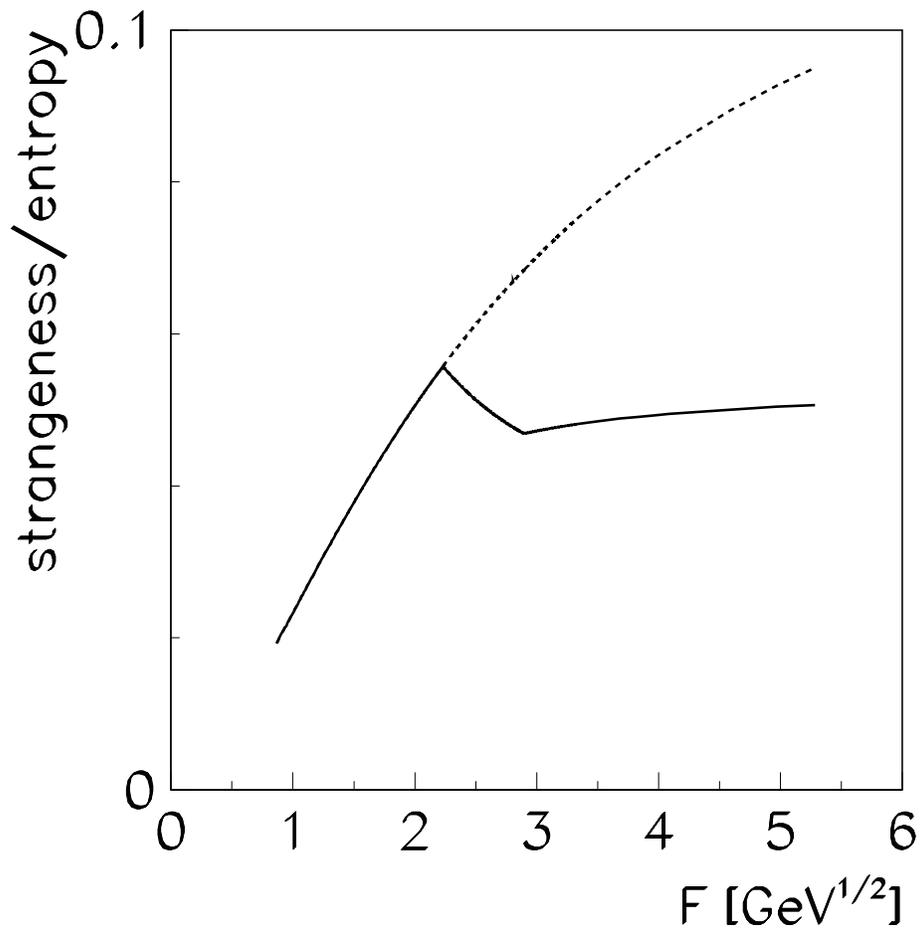,width=14cm}
\caption{
The ratio of the total number of $s$ and $\overline{s}$ quarks and
antiquarks to the entropy (solid line) as a function of $F$.
The dashed line indicates the corresponding ratio calculated
assuming absence of the phase transition to the QGP.
}
\label{str}
\end{figure}

We are interested in the collision energy region between
the AGS and SPS.
At `low' collision energies (when a pure W--state is formed)
the strangeness to entropy ratio increases with $F$.
This is due to the fact
that the mass of the strange degrees of freedom is significantly 
higher than the system temperature.
At $T = T_c$ the ratio is higher in the W--state than in the  Q--state, this
causes the decrease  
of the ratio in the mixed phase to the level characteristic for the Q--state.
In the Q--state, due to the low mass of strange quarks  in comparison
to the system temperature, 
only a weak dependence
of the ratio on $F$ is observed.
The $F$ dependence of strangeness/entropy ratio is shown in
Fig.~\ref{str}.

Within the model one can estimate the lower limit for
$T_c$ assuming that the transition 
starts just above top AGS energy (15 A$\cdot$GeV).
In this case one obtaines $T_c = 170$
($B$ = 300 MeV/fm$^3$) and the non--monotonic behaviour
of the strangeness production is substituted by a rapid
saturation.
Remaining signatures of the phase transition are unchanged.

\newpage

\section{Comparison with Data}

The comparison of the model with experimental data
on pion and strangeness production is presented below.
The results are taken from the compilations
\cite{Ga:95, Ga:96, Ga:97} where the references to the
original experimental publications can be found. 

During the evolution of the system the equilibration between
newly created matter and baryons takes place.
It is argued that this equilibration causes transfer of 
entropy from the produced matter to baryons.
The analysis of the pion suppression effect at low collision energies
indicates that this transfer corresponds to the effective
absorption of about 0.35 pion per participant nucleon \cite{Ga:96a}.
We assume that there are no other processes which change the entropy
content of the state produced in the early stage.

For the comparison with the model it is convenient  to define
the quantity:
\begin{equation}
\langle S_{\pi} \rangle = \langle \pi \rangle + \kappa
\langle K + \overline{K} \rangle + \alpha \langle N_P \rangle, 
\end{equation}
where $\langle \pi \rangle$ is the measured total multiplicity of
final state pions and $\langle K + \overline{K} \rangle$
is the multiplicity of kaons and antikaons.
The  factor $\kappa$ = 1.6 is the approximate 
 ratio between mean entropy carried by a single kaon to the 
corresponding pion entropy at chemical freeze--out. 
The term $\alpha \langle N_P \rangle$ with $\alpha$ = 0.35
is a correction for the above discussed transfer of the entropy to
baryons.
The quantity $\langle S_{\pi} \rangle$ can thus be  interpreted 
as the early stage entropy measured in  pion entropy units.
The conversion factor between $S$ and $\langle S_{\pi} \rangle$
is choosen to be 4 ($\approx$ pion entropy at chemical freeze--out).

The number of baryons which take part in the collision ($2A_p$ in the model
calculations) is identified now with the experimentally measured number of
participant nucleons, $\langle N_P \rangle$.
The fraction of energy carried by the produced particles 
($\eta$ in Eq. 1) is taken to be 0.67 as measured by the NA35  
Collaboration \cite{Ba:94} for central S+S collisions at 200 A$\cdot$GeV.
Production of pions and kaons scales with the number of participant nucleons
when central Pb+Pb and S+S collisions at SPS are compared \cite{Af:96}.
This suggests that $\eta$ can be assumed to be independent of the size
of the colliding nuclei.
Similar values of $\eta$ are obtained when  central A+A collisions at
the AGS are analyzed \cite{St:96} and the correction for the pion absorption
is taken into account.

\begin{figure}[p]
\epsfig{file=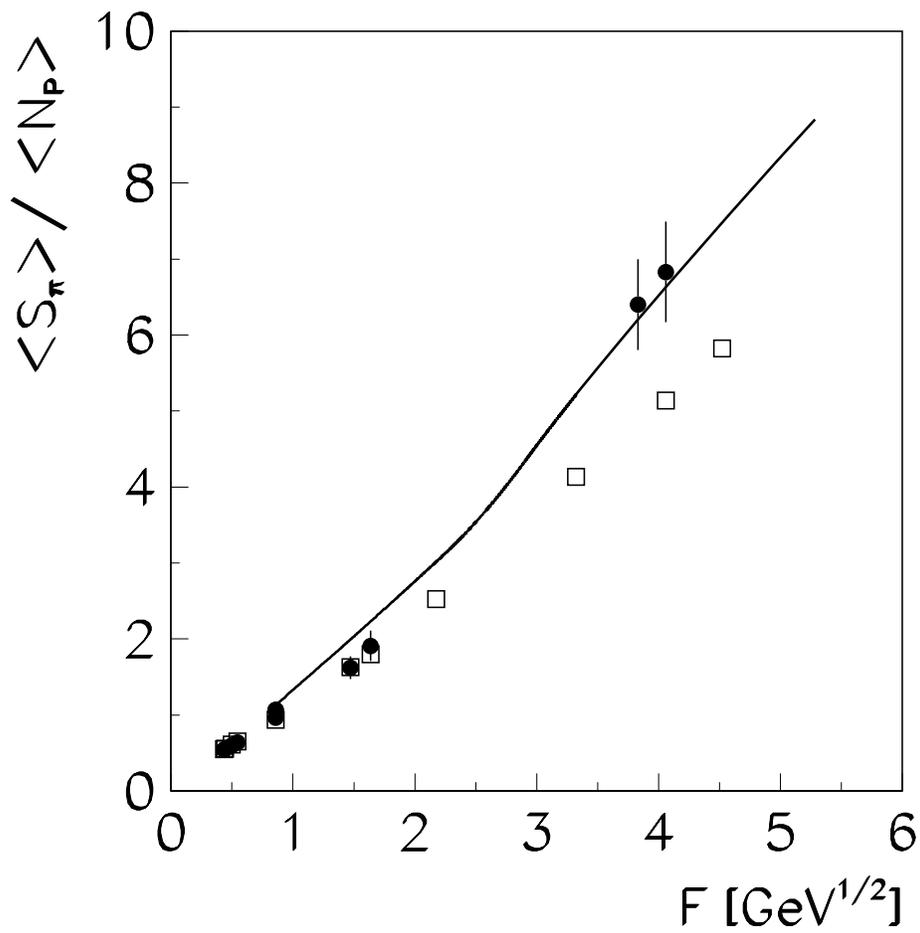,width=14cm}
\caption{
The $\langle S_{\pi} \rangle/\langle N_P \rangle$ ratio as a function
$F$. Experimental data on central collisions of two identical
nuclei are indicated by  closed circles.
These data should be compared with the model predictions shown  by 
the solid line.
The open boxes show results obtained for nucleon--nucleon interactions.
}
\label{pipb}
\end{figure}

The comparison between  data on
$\langle S_{\pi} \rangle/\langle N_P \rangle$ and the model is shown
in Fig. \ref{pipb}.
The parametrization of the W--state has been chosen
to fit the AGS data and, therefore, an agreement with
low energy A+A data is not surprising. 
On the other hand the description of  high energy (SPS) results
obtained by the NA35 and NA49 Collaborations is essentially
parameter free, as the properties of the early stage state,
Quark Gluon Plasma, are rather well defined.
The change of the slope in $F$ dependence of the pion multiplicity
was previously proposed as a signature of the transition region
\cite{Ga:95a}.

The comparison between the model and the data on strangeness
production is done under the assumption that the strangeness content
defined in the early stage is preserved till the hadronic freeze--out.
It simplifies the  picture 
as the gluon contribution to the strangeness production
during the QGP hadronization is neglected.
We do not expect a significant number of strange
($s,\bar{s}$)-pairs or/and strange-antistrange hadron pairs
produced by massless gluons with 
typical momenta of several hundred MeV at $T=T_c$.
   
\begin{figure}[p]
\epsfig{file=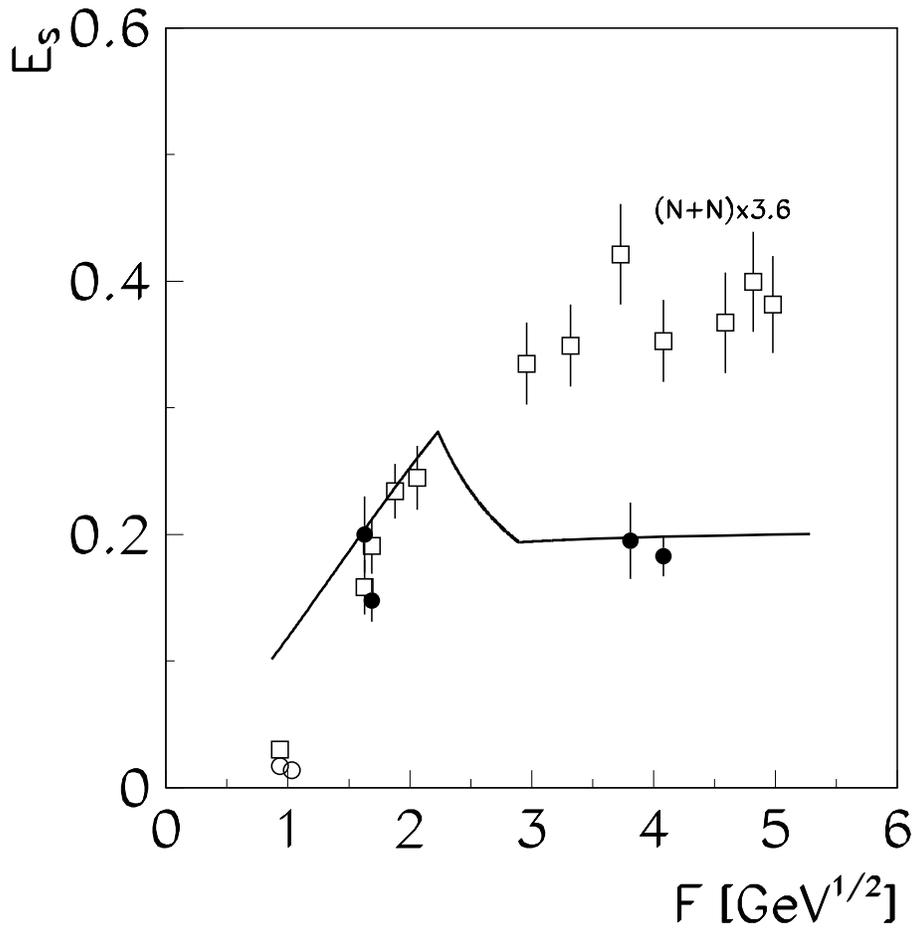,width=14cm}
\caption{
The ratio $E_S$ as a function
$F$. Experimental data on central collisions of two identical
nuclei are indicated by  closed circles.
These data should be compare with the model predictions shown  by 
the solid line.
The open boxes show results obatined for nucleon--nucleon interaction,
scaled be a factor 3.6 to match A+A data at AGS energy.
}
\label{es}
\end{figure}

The total strangeness production is usually studied using
the experimental ratio:
\begin{equation}\label{esexp}
E_s ~=~ \frac {\langle \Lambda \rangle + \langle K + \overline{K} \rangle}
            {\langle \pi \rangle},
\end{equation}
where $\langle \Lambda \rangle$ is the mean multiplicity of $\Lambda$
hyperons.
Within the model $E_s$ (\ref{esexp}) is calculated as:
\begin{equation}\label{esmodel}
E_s ~=~ \frac {(N_s + N_{\overline{s}})/\zeta}
            { (S - S_s)/4 - \alpha \langle N_P \rangle },
\end{equation}
where  $\zeta$ = 1.36 is the experimentally estimated ratio between
total strangeness and strangeness carried by  $\Lambda$ hyperons and
$K$ + $\overline{K}$ mesons \cite{Bi:92} and          
$S_s$ is the fraction of the entropy carried by the strangeness
carriers.
The comparison between the calculations and the data is shown in
Fig. \ref{es}.
The description of the AGS data is again a consequence
of our parametrization of W-state: $g_W^s = 14$, $m_W^s = 500$ MeV. 
As in the case of the pion multiplicity, the description of the
strangeness results at the SPS (NA35 and NA49 Collaborations)
can be considered as being essentially parameter free
\footnote{ The $E_s$ value resulting from a QGP can be estimated in a simple 
way. Assuming that $m_s = 0$, and neglecting the small ($< 5 \%$) effect
of pion absorption at the SPS, 
one gets from  (\ref{strent}) and (\ref{esmodel}) 
$E_s \approx (g^s_Q/1.36)/g^{ns}_Q \approx 0.21$,
where $g^s_Q = (7/8) \cdot 12$ 
is the effective number of degrees of freedom of
$s$ and $\overline{s}$ quarks and $g^{ns}_Q = 16 + (7/8) \cdot 24$ 
is the corresponding number for $u, \overline{u}, d, \overline{d}$
quarks and gluons. Here we also use  the approximation that the
pion entropy at  freeze--out is equal to the mean
entropy of $q$, $\overline{q}$ and $g$ in a QGP.}.
The agreement with the SPS data is obtained assuming creation of 
globally equilibrated QGP in the early stage of nucleus--nucleus collisions.
The characteristic non--monotonic energy dependence of the $E_S$ ratio
was proposed in Ref. \cite{Ga:96} as a signature of the phase transition 
and it is confirmed
here by  calculations in our model.
Measurements of strangeness and pion production in the transition region
are obviously needed.

\begin{figure}[p]
\epsfig{file=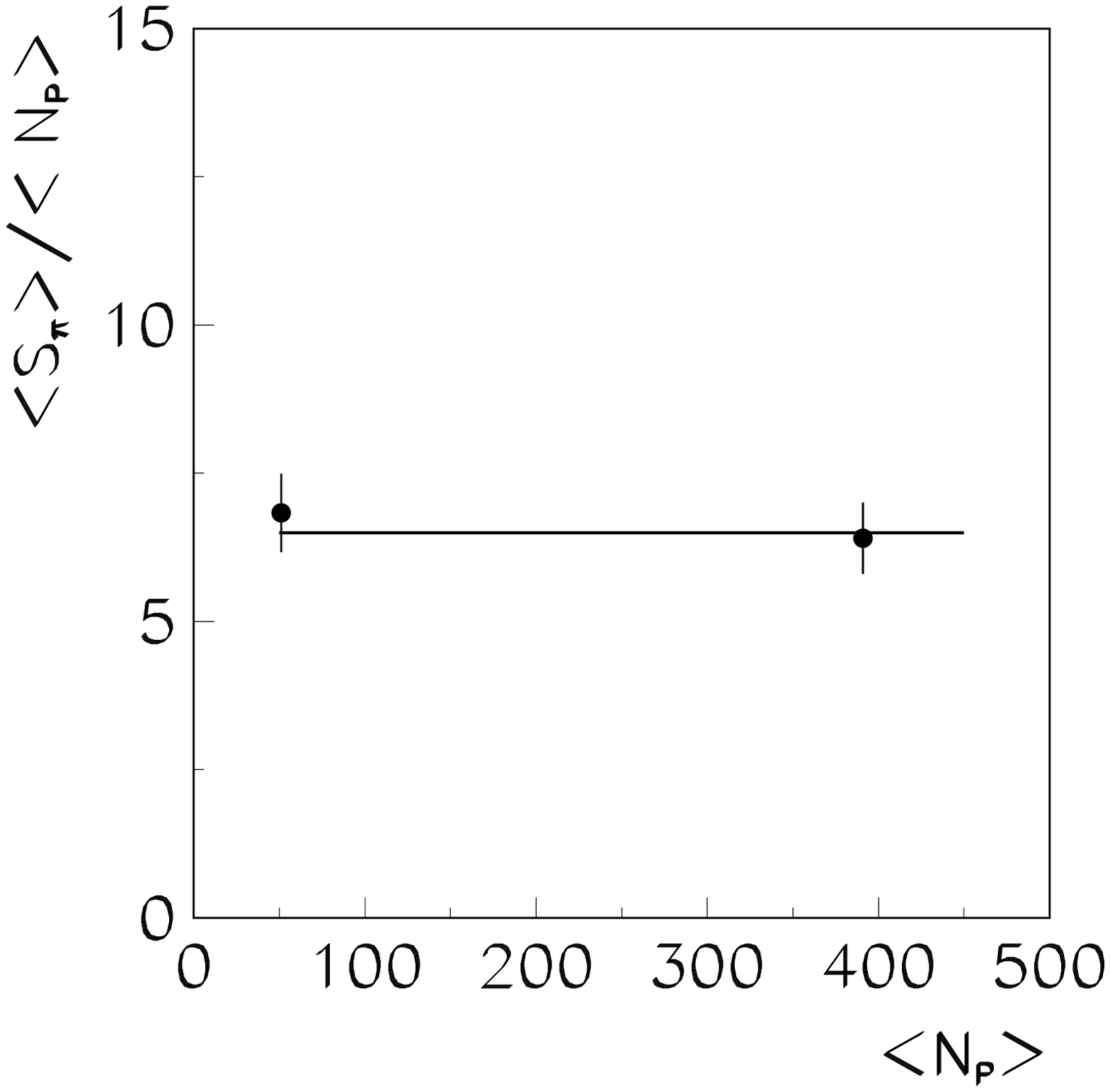,width=14cm}
\caption{
The $\langle S_{\pi} \rangle/\langle N_P \rangle$ ratio as a function
$\langle N_P \rangle$
for central S+S and Pb+Pb collisions at 200 A$\cdot$GeV and 158 A$\cdot$GeV.
The results are not corrected for a small difference in the
collision energy (see Fig. \protect\ref{pipb}). 
The model prediction is shown  by 
the solid line.
}
\label{pipb_vs_np}
\end{figure}

\begin{figure}[p]
\epsfig{file=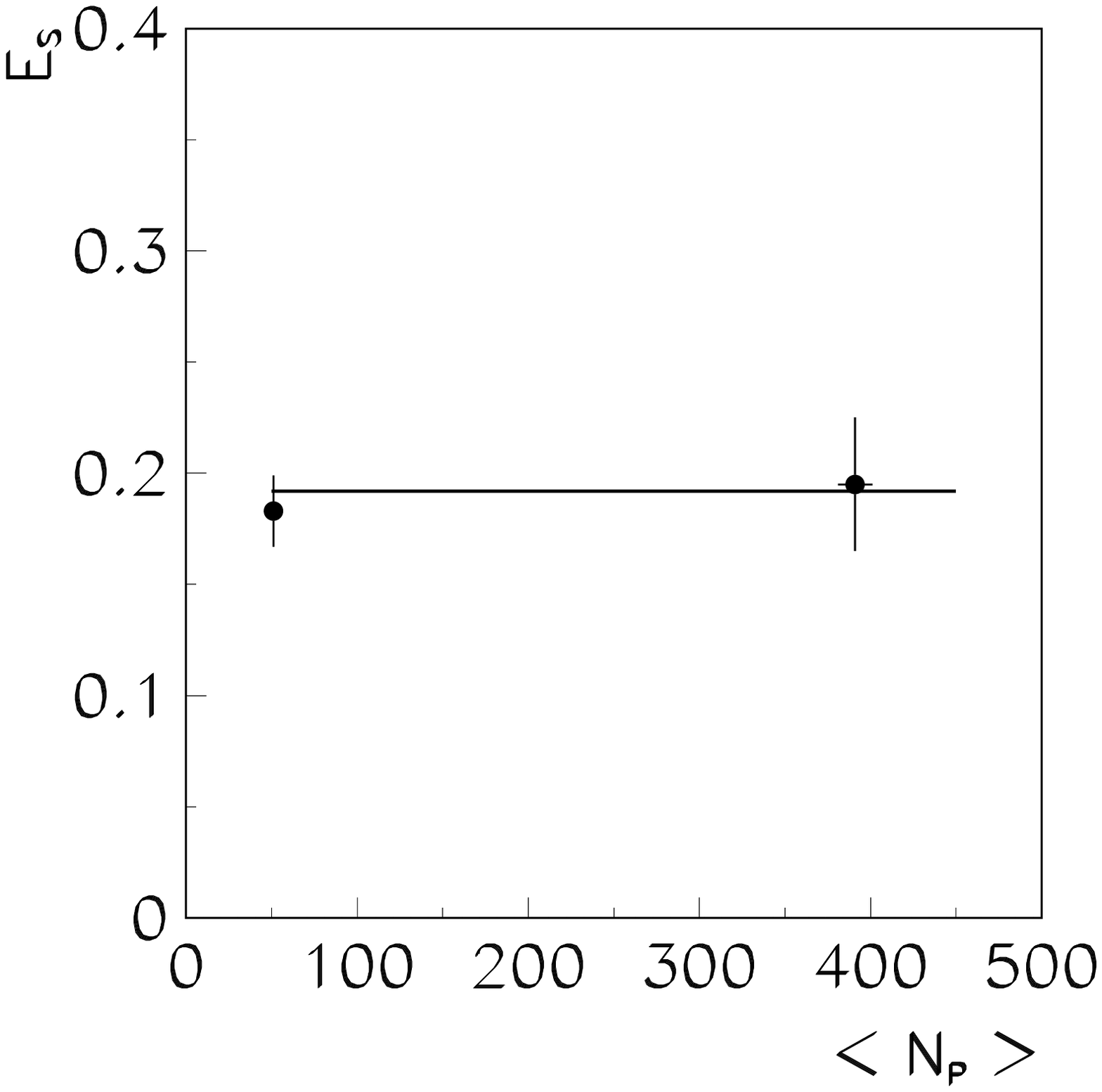,width=14cm}
\caption{
The ratio $E_S$ as a function
$\langle N_P \rangle$
for central S+S and Pb+Pb collisions at 200 A$\cdot$GeV and 158 A$\cdot$GeV.
The results are not corrected for a small difference in the
collision energy (see Fig. \protect\ref{es}).
The model prediction is shown  by
the solid line.
}
\label{es_vs_np}
\end{figure}

The entropy and strangeness production in central A+A collisions
considered here satisfies well the conditions needed for thermodynamical
treatment.
Therefore one expects that the measures of the entropy per participant
nucleon, 
$\langle S_{\pi} \rangle/\langle N_P \rangle$, and strangeness per
entropy, $E_s$, are independent of the number of participants
for large enough values of $\langle N_P \rangle$.
In order to check this in an explicit way we show 
$\langle S_{\pi} \rangle/\langle N_P \rangle$ (Fig. \ref{pipb_vs_np})
and $E_s$ (Fig. \ref{es_vs_np}) as a function of $\langle N_P \rangle$
at SPS energy for central S+S and Pb+Pb collisions.

\newpage

\section{Discussion}

The relationship between our approach and two widely discussed aspects
of nucleus--nucleus collisions namely strangeness  and $J/\psi$
production are presented.
Finally we comment on the event--by--event fluctuations.

\subsection{Strangeness Production}

The enhanced production of 
strangeness was considered by many authors as a potential
signal of QGP formation \cite{Ko:86, Ka:86, Ma:86}.
The line of arguments is the following. 
One estimates that the strangeness equilibration time in QGP is 
comparable to the duration of the collision process ($< 10 $ fm/c)
and about 10 times shorter than the corresponding equilibration time in
hadronic matter.
It is further assumed that in the early stage the strangeness density is much
below the equilibrium density e.g. it is given by the strangeness
obtained from the superposition of nucleon--nucleon interactions.
Thus it follows that during the expansion of the matter the strangeness
content increases rapidly and approaches its equilibrium value
provided  matter is in the QGP state.
In the case of hadronic matter the modification of the initial strangeness
content is less significant due to the long equilibration time.
This leads to the expectation that strangeness production should 
rapidly increase when the energy transition region is crossed from below.

In the model presented in this paper the role of strangeness is  different.
The reason can be found in the  assumption
concerning 
the early stage properties.
We assume that due to the statistical nature
of the  creation process
the strangeness in the
early stage is already in equilibrium and therefore possible secondary
processes do not modify its value. 
As at $T = T_c$ the strangeness density is similar or even lower 
(depending on the $T_c$ value) in the 
QGP than in
the confined matter, 
saturation or suppression of strangeness production is expcted to
occur when crossing the transition energy range from below.

In our model the low level of strangeness production in N+N interactions
as compared to the strangeness yield in central A+A collisions,
called strangeness enhancement, 
can be understood as due to the effect of strict strangeness conservation
(canonical suppression factor) imposed on the degrees of freedom in the confined
matter in the early stage.

\subsection{$J/\psi$ Production}

Suppression of $J/\psi$ production was proposed as a signal of creation of
the QGP in nuclear collisions \cite{Ma:86a}.
The details of the  models used to describe this process changed with time, 
but the 
main line of arguments of relevance here is the same since the first
proposal (for a recent review see \cite{Ge:98}).
The interpretation of $J/\psi$ results is done within 
a hard production QCD model.
It is assumed that the creation of $J/\psi$  follows
the dependence given by the production of Drell--Yan pairs,
i.e. the inclusive cross section in A+A collisions
increases with $A$ as $A^2$.
Deviations from this dependence  are interpreted as due to interactions of
the $J/\psi$ (or `pre--$J/\psi$' state) with the surrounding matter.
Suppression of $J/\psi$ observed in p+A and O(S)+A collisions at SPS
is considered to be caused by 
the interactions with  participant nucleons and
produced particles.
The rapid increase of the suppression  observed only for central
Pb+Pb collisions is attributed 
\cite{Bl:97, Kh:97, Sa:97}  to the formation of 
a QGP which leads
to a strong additional desintegration of the  $J/\psi$ mesons. 

This interpretation is in  contradiction to the conclusions based on the
analysis of pion and strangeness results within the  statistical model.
No anomalous change in this observables  is seen between central S+S 
and cental Pb+Pb collisions 
(see Figs. \ref{pipb_vs_np} and \ref{es_vs_np}).
It is therefore essential to understand whether this contradiction can be
removed when the same approach is used to interpret the whole set of 
data.

It is natural to extend our statistical model to charm production
assuming that like entropy and strangeness, 
charm in the early stage is produced according to  phase space.
We take the mass of the charm
quark to be $m_Q^c \cong$ 1.5~GeV and calculate the mean number of $c$ and
$\overline{c}$ quarks for central Pb+Pb collisions at 158 A$\cdot$GeV.
The early stage volume (\ref{volume}) for the experimental number
of participant nucleons  $\langle N_P \rangle$
in central Pb+Pb collisions is approximately $V\cong$ 200~fm$^3$
and the early stage temperature is 264 MeV (see Fig. ~\ref{temp}).
For the number density of charm quarks and antiquarks
($g_Q^c=2\cdot 2\cdot 3=12$) we get:
\begin{equation}\label{cantic}
\rho _{c}~=~\frac{g_Q^c}{2\pi^2} 
\int_{0}^{\infty}k^2dk~\frac{1}{\exp\left(\frac {\sqrt{k^2+(m_Q^c)^2}}{T}
\right)~+~1} ~
\cong~g_Q^c \left(\frac{m_Q^c T}{2\pi}\right)^{3/2}
\exp\left(-~\frac{m_Q^c}{T}\right)~. 
\end{equation}
Finally  the total average number of charm quarks
and antiquarks can be estimated as:
\begin{equation}\label{nc}
N_{c}~=~\rho_{c}~V~ \cong~0.085~\mbox{fm}^{-3}\times 200~
\mbox{fm}^3~ \cong~ 17~.
\end{equation}
Note that the contribution of $c$ and $\overline{c}$ to the thermodynamical
functions of the QGP was neglected in the calculations presented in the 
previous
Sections. 
This is indeed justified by a large value of the charm quark mass.
The contribution of charm quarks and antiquarks to the energy density can be 
estimated as $\varepsilon_c\cong \rho_c m_Q^c\cong$ 0.13 GeV/fm$^3$.
This value is much smaller than the total energy density of the QGP,
$\varepsilon_Q\cong$11~GeV/fm$^3$,
in the early stage of Pb+Pb at the SPS. 
The inclusion of charm
into the QGP equation of state causes a decrease of  less than 1~MeV
of the early stage temperature.  

The equilibrium number (\ref{nc}) 
exceeds substantially
the  estimate given in Ref.~\cite{Br:97} which  is based on the assumption of
perturbative production of open charm.
Recently the NA50 Collaboration attempted to estimate open charm
production in central Pb+Pb collisions at the SPS using the measured
invariant mass spectrum of dimuon pairs \cite{Sc:97}.
This estimate relies on the assumption 
(based on the PHYTIA model) that the production of $c$ and $\overline{c}$ quarks
is correlated.
In our statistical model $c$ and $\overline{c}$ quarks are independent
and therefore their contribution to the dimuon spectrum
can not be distinguished from the background 
contribution\footnote{We thank E. Scomparin for pointing to us this
property of the NA50 procedure.}.
Thus results of NA50 on `charm--like' enhancement can not be
compared with our predictions.

\begin{figure}[p]
\epsfig{file=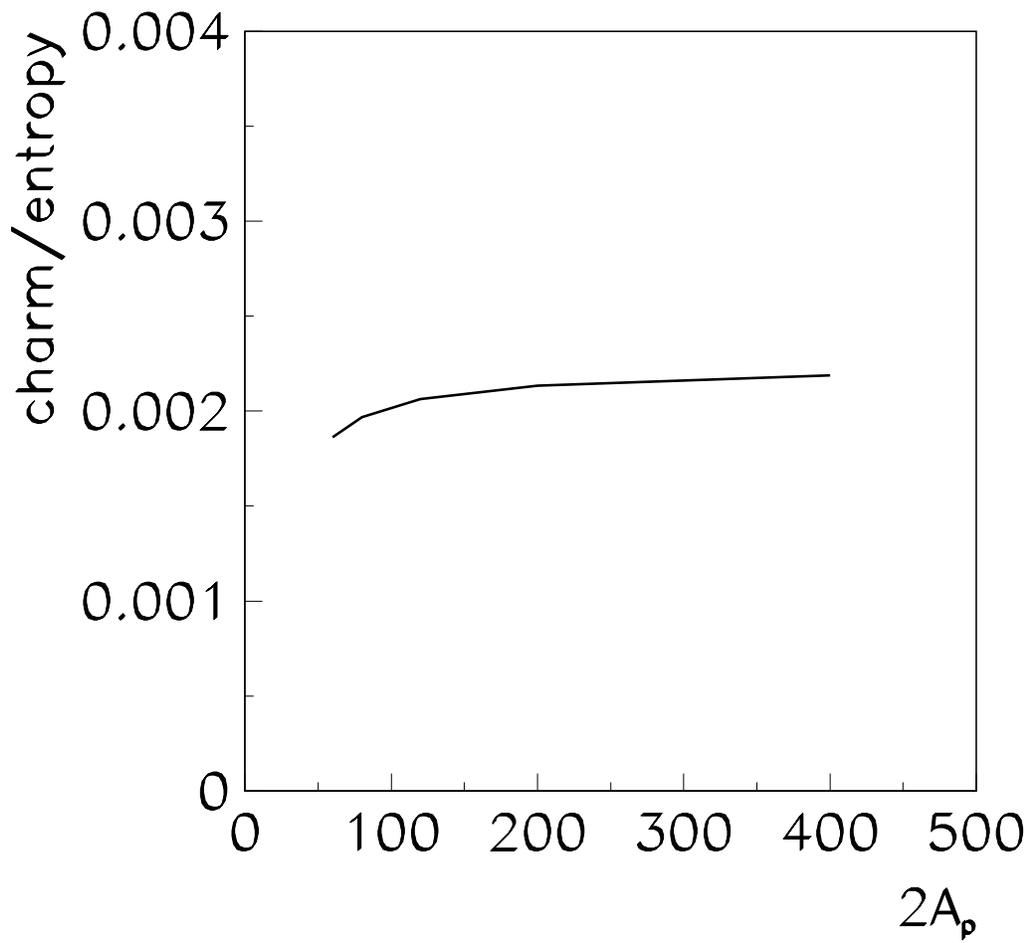,width=14cm}
\caption{
The ratio of charm to entropy as a function of the
the number of participant nucleons ($2A_p$) for A+A collisions at
158 A$\cdot$GeV.
The canonical suppression factor is included in the calculation.
}
\label{cps}
\end{figure}

Due to the large mass of the charm quark one should
consider a possible correction to the grand canonical
approximation due to strict
charm conservation.
In Fig. \ref{cps} we show the predicted dependence of the 
charm to entropy ratio
on the number of participant nucleons ($2A_p$) including a correction for
strict charm conservation calculated as in Ref.~\cite{Ra:80}.
It is observed
that even down to low values of $2A_p = 40$ 
the correction is small and therefore the charm/entropy
ratio is approximately independent of the volume of the system,
similar to the strangeness/entropy ratio.

\begin{figure}[p]
\epsfig{file=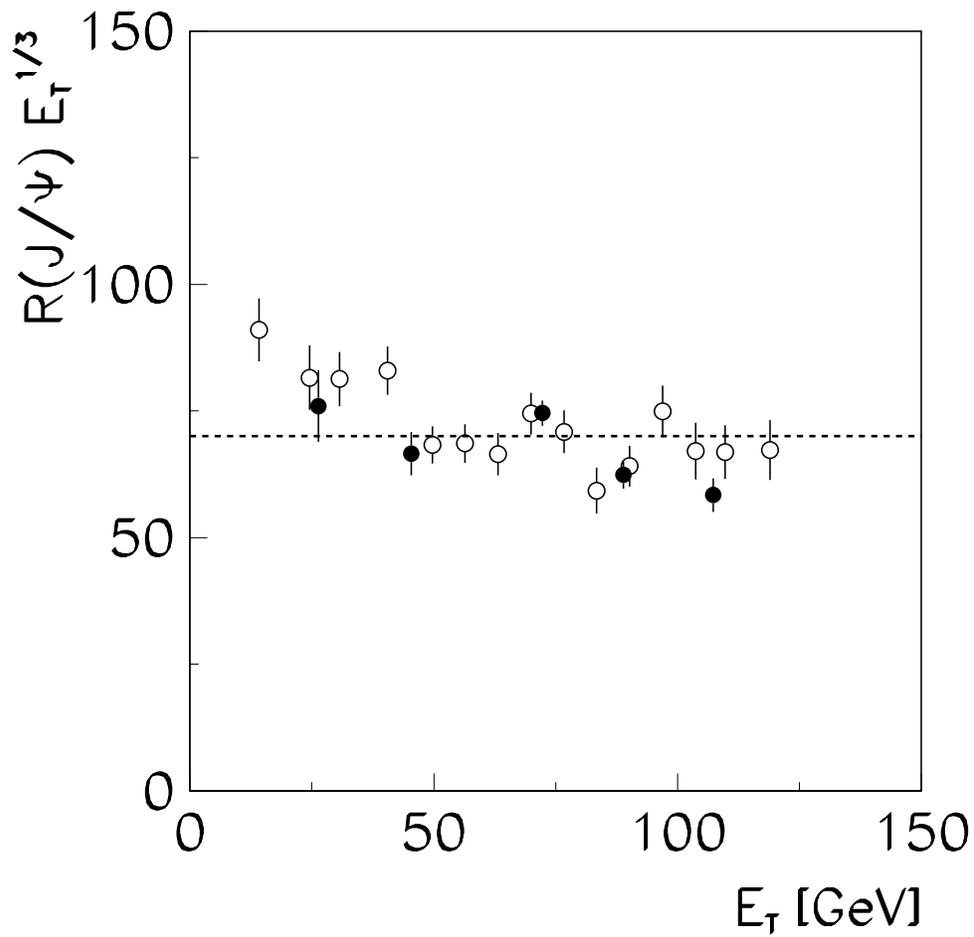,width=14cm}
\caption{
The estimate of the 
transverse energy dependence of the
$J/\psi$ multiplicity per pion in
collisions at 158 A$\cdot$GeV.
The closed points show final 1995 data and the open points preliminary 1996 data of 
the NA50 Collaboration.
The dashed line is drawn for the reference.
}
\label{psippi}
\end{figure}

Analysis of the hadron yields within a statistical 
hadronization model \cite{Be:96, Be:97}
shows that hadronization is a local statistical process.
Thus one expects that also the ratio of the  mean $J/\psi$
multiplicity to entropy (pion multiplicity) should be volume
independent.
This prediction of our model can be  checked against experimental data.
However as the results on $J/\psi$ multiplicity are not published,
we have to perform ourselves a conversion of the
available data.
We start from  the $E_T$ dependence of the ratio measured by
the NA50 Collaboration \cite{Ra:97}:
\begin{equation}
R(J/\psi) = \frac {B_{\mu\mu} \sigma(J/\psi)}  {\sigma(DY)}~,
\end{equation}
where $\sigma(J/\psi)$ and $\sigma(DY)$ are inclusive cross sections
for production of $J/\psi$ and Drell--Yan pairs and
$B_{\mu\mu}$ is the branching ratio for $J/\psi$ decay into a $\mu^+ \mu^-$ 
pair.
As at SPS energies pion production dominates particle production, the
measured transverse energy is basically determined 
by the pion transverse energy.
The mean transverse momentum of pions is indepedent of the 
centrality \cite{Al:97} 
and
therefore
$E_T$ can be considered to be proportinal to the pion
multiplicity or the number of participant nucleons.
The multiplicity of Drell--Yan pairs increases with the
centrality as $\langle N_P \rangle^{4/3}$ or equivalently as $E_T^{4/3}$.
Thus the mean $J/\psi$ multiplicity 
is expected to increase  as:
\begin{equation}
\langle J/\psi \rangle \sim R(J/\psi) E_T^{4/3}
\end{equation}
and consequently the $J/\psi$ multiplicity per pion should be proportional to
\begin{equation}
\frac {\langle J/\psi \rangle} {E_T}  \sim R(J/\psi) E_T^{1/3}.
\end{equation}
The values of  $R(J/\psi) E_T^{1/3}$ are plotted as a function of $E_T$ 
for Pb+Pb collisions at 158 A$\cdot$GeV in Fig.~\ref{psippi}.
The ratio 
 $R(J/\psi) E_T^{1/3}$
($\sim \langle J/\psi \rangle/\langle \pi \rangle$) seems to be 
independent of $E_T$  in the whole
range of $E_T$.
Thus we conclude that the experimental dependence of $J/\psi$ production
on $E_T$ in Pb+Pb collisions is in the  agreement with the expectation
of our model.
It just reflects the statistical character of charm production and the following
hadronization process.

As pointed out in Ref. \cite{Be:97}
the particle abundances resulting from   hadronization  can be
modified by inelastic interactions in the freezeing--out hadronic
matter.
It is however argued \cite{Ge:98}
that the $J/\psi$ hadronic cross sections are small, thus
no significant reduction of the $J/\psi$ yield is expected due to 
hadronic interactions.
This argumentation is however not valid for $\psi'$ production for which
hadronic cross sections are estimated to be 10 times larger 
than for $J/\psi$ and therefore
a significant suppression of  $\psi'$ mesons 
can be  observed \cite{Ge:98}.
This is usually considered as the reason why the ratio $\sigma(\psi')/\sigma(J/\psi)$ 
decreases with
$E_T$  \cite{Ge:98}.

We summarize that  the  effect of the anomalous $J/\psi$ suppression
in central Pb+Pb collisions is a result  of the interpretation of the
data within  
the model of hard  QCD production of $J/\psi$ with 
following suppression.
The same data analyzed within the statistical approach show no
anomalous behaviour and lead to a consistent interpretation
of the results on pion, strangeness and charm production.

\subsection{Event--by--Event Fluctuations}

It was recently measured by the NA49 Collaboration
\cite{Ro:97} that event--by--event transverse momentum fluctuations
in central Pb+Pb collisions at 158 A$\cdot$GeV are  smaller
than  fluctuations measured in  p+p interactions and 
expected in non--equilibrium
models of nuclear collisions \cite{Ga:92, Ga:97a}.
A decrease of the global fluctuations with the increasing volume of the system
and/or increasing number of internal degrees of freedom 
is a generic feature of statistical models \cite{St:95}.
Therefore  in our  approach one expects 
a decrese of global fluctuations when going from p+p interactions
to central Pb+Pb collisions.
The same arguments lead to the conclusion that the flavour fluctuations
should be also reduced in central A+A collisions in comparison to p+p
interactions.
The method to analyze these fluctuations was recently formulated \cite{Ga:98}.

Finally we expect an increase of the fluctuations in the transition region.
This is because of the additional possibility of changing the
relative content of W-- and Q--state in the early stage. 
An important observable should be the strangeness to entropy ratio as
it is significantly different in W-- and Q--state at  $T = T_c$.

\newpage
\section{Summary and Conclusions}

In this paper we further 
develop  the statistical model of the early stage
in high energy nucleus--nucleus collisions.
We attempt to understand the possible meaning of the equilibration
of the {\it created state}. 
We attribute the success of the statistical description of the early
stage of A+A collisions to the statistical nature of primary creation process   
rather than to the
result
of following multiple secondary interactions.

We show that the assumption that the state created in the
early stage  is in the form of
a Quark Gluon Plasma gives an essentially parameter free description of the
data on pion and strangeness production in central A+A collisions at SPS.

It is argued that the early stage  degrees of freedom 
in confined matter
can not be modeled by  hadrons and hadronic resonances.
An effective statistical description of the confined state (W--state) 
is introduced and the 
parameters  characterizing degrees of freedom
are extracted from comparison to data.

The transition between W--state and QGP when increasing collision energy is
discussed.
It is proven that the condition of maximum entropy is equivalent to
the Gibbs
construction of the first order phase transition between W--state and QGP.

The transition region is localized to be between 30 A$\cdot$GeV and 
65 A$\cdot$GeV
for the set of parameters used in the paper.
It is shown that the transition should be associated 
with a  characteristic 
increase  of pion multiplicity and a non--monotonic energy dependence 
of the strangeness to  pion ratio.
It is also argued that an increase of the  event--by--event fluctuations
can be expected
in the transition region.
Note that
anomalies in the  space--time pattern of the matter expansion are also 
expected
due to softening of the equation of state in the mixed phase \cite{Hu:95}.
This can be detected by the analysis of single particle spectra and two
particle correlations \cite{Le:90}.

Finally we remind
that  the anomalous $J/\psi$ suppression in central Pb+Pb
collisions at the SPS is a result of the data interpretation
within a model assuming that the charm production is a hard QCD process. 
We show that the same experimental  
results are also consistent with the hypothesis that the
$J/\psi$ multiplicity  per pion is independent of the centrality of Pb+Pb
collisions, similar to the behaviour of the strangeness/pion ratio.
This behaviour can be reproduced in our 
approach when the charm production is treated 
in the same statistical way as the production of strangeness.
It allows for a consistent interpretation of the results
on pion, strangeness and $J/\psi$ production in A+A collisions at SPS.
Data on total charm production are obviously needed to check our assumption
of a statistical nature of the charm production.

We conclude that a broad set 
of experimental data is in agreement with the hypothesis
that a QGP is created in central A+A (S+S and Pb+Pb) collisions at the SPS.
A study of the energy dependence of several basic observables (pion and strangeness
multiplicities, expansion pattern 
and event--by--event fluctuations) should be able to uniquely
prove the existence of a 
 phase transition to a Quark Gluon Plasma.

\newpage

\noindent{\Large \bf Appendix A}\\

The properties of the W--state were obtained by an `educated guess'
procedure. It is however still intresting to note that the degrees of
freedom in  W--state can be identified with  colourless
$q\overline{q}$ pairs. Assuming that the light quarks ($u$, $d$) are almost
massless we obtaine 4 nonstrange flavour--antiflavour combinations:
$u\overline{u}$, $d\overline{d}$, $d\overline{u}$ and $u\overline{d}$.
Each pair can be in 4 spin states, which gives 16 massless nonstrange
internal degrees of freedom. Similar counting gives 16 different pairs
with $s$ or $\overline{s}$ quark. There are in addition 4
$s\overline{s}$ pairs which however can be considered strongly suppressed
due to the large mass of strange quarks. 
Thus the numbers of non--strange and strange degrees of freedom
obtained for the colourless $q \overline{q}$ pairs
approximately coincide with the corresponding numbers extracted
from the data for the W--state ($g_W^{ns} \approx 16$ and
$g_W^{s} \approx 14$).

The reduction of the effective number of degrees of freedom
from colored $q$, $\overline{q}$ and $g$ to colour neutral
$q\overline{q}$ pairs may be understood as a result of the 
requirement of local colour neutrality imposed on the creation
process at low energy density and/or for small systems.

We note also that the colourless $q \overline{q}$ pairs
are identified as  initial degrees of freedom in the
$e^++e^-$ anihilation process
by the well known analysis of the ratio
$\sigma(e^++e^- \rightarrow hadrons)$/$\sigma(e^++e^- 
\rightarrow \mu^+ + \mu^-)$ \cite{Pe:82}.

\vspace{0.5cm}
\noindent 
{ \Large \bf Appendix B}\\ 

We present here a general proof of the equivalence between
Gibbs construction of the 1st order phase transition and the
basic condition that the equilibrium state is equal to the maximum
entropy state.
The proof is valid  when all conserved charges are
equal to zero as considered in the paper.
In this case the 
pressure function, $p=p(T)$,
defines completely the system thermodynamics 
provid that the system volume $V$
goes to infinity (thermodynamical limit).
 The temperature $T$ remains the
only independent thermodynamical variable.  The energy density,
$\varepsilon (T)$, and entropy density, $s(T)$, are calculated as
\begin{equation}\label{epsands}
\varepsilon (T)~ =~T\frac{dp}{dT}-p~,~~~~
s(T)~=~\frac{dp}{dT}~.
\end{equation}

A discontinuity of the first derivative,
$dp/dT$, at  $T=T_c$ corresponds, by definition, to a 1st order
phase transition at temperature $T=T_c$.  In physical terms one describes
the system at $T<T_c$ by a function $p=p_1(T)$ (low--temperature phase) and
by $p=p_2(T)$ at $T>T_c$ (high--temperature phase).  At $T=T_c$ the pressures
of the two phases are equal 
\begin{equation}\label{preseq}
p_1(T_c)~=~p_2(T_c)~\equiv p_c~, 
\end{equation} 
and their first derivatives
satisfy the inequality 
\begin{equation}\label{presdir}
\left(\frac{dp_2}{dT}\right) _{T=T_c}~>~
\left(\frac{dp_1}{dT}\right)_{T=T_c}~.  
\end{equation} 
The energy density
discontinuity (latent heat) as well as the entropy density discontinuity
take place at $T=T_c$:  
\begin{equation}\label{lheat1} 
\Delta \varepsilon
~=~\varepsilon _2(T_c)- \varepsilon _1 (T_c)~= 
~ T_c ~[s_2(T_c)-s_1(T_c)]~>~0~.  
\end{equation} 
At $T=T_c$ the system is in the
mixed phase with 
\begin{equation}\label{mixedphase} 
\varepsilon_{mix}~ =~
(1-\xi)\varepsilon_1(T_c)~+~\xi \varepsilon_2(T_c)~, ~~~
s_{mix}~=~(1-\xi)s_1(T_c)~+~\xi s_2(T_c)~, 
\end{equation} 
where $1-\xi$ and
$\xi$ are relative volumes occupied by phases `1' and `2',
respectively. The above construction is known as the Gibbs creteria for a
1st order phase transition: at a given temperature $T$ the system occupies
a pure phase whos pressure is larger. 
The mixed phase is formed if
both pressures are equal. One considers phases `1' at $T>T_c$ and `2'
at $T<T_c$ as metastable states (superheated and supercooled,
respectively). Such a consideration is physically important in the kinetic
picture of a phase transition and for the studies of statistical
fluctuations.   We prove now the
equivalence of the Gibbs creteria to the maximum entropy condition of the
mixed phase.  It claims that at any energy density $\varepsilon$ from the
interval $[\varepsilon_1(T_c),\varepsilon_2(T_c)]$ the entropy density of
the mixed phase is maximal:  
\begin{equation}\label{mixextrem}
s_{mix}(\varepsilon)~>~s_i(\varepsilon)~,~~~ i=1,2~,~~~ \varepsilon \in
[\varepsilon_1(T_c),\varepsilon_2(T_c)]~.  
\end{equation} 
The following equations for the entropy densities of
the pure and mixed phase can be easily obtained from Eq.~(\ref{epsands}): 
\begin{equation}\label{si} 
s_i~=~\frac{\varepsilon + p_i(T)}{T}~,~~
i=1,2~;~~~~ s_{mix}~=~\frac{\varepsilon + p_c}{T_c}~.  
\end{equation}
Now
the values of $s_1$ and $s_2$ should be compared to the value of $s_{mix}$ 
at the same
$\varepsilon$ from the interval $[\varepsilon_1(T_c),\varepsilon_2(T_c)]$. 
This means that the comparison is done at the temperature of a
pure phase
$T>T_c$ for $i$=1 and $T<T_c$ for $i$=2 in Eq.~(\ref{si}). 
The inequalities
(\ref{mixextrem}) can be  transformed into 
\begin{eqnarray}\label{ineq}
\frac {dp_1}{dT}~&>&~\frac{p_1(T)-p_c}{T-T_c}~,~~~~T>T_c~,\\ 
\frac{dp_2}{dT}~&<&~\frac{p_c-p_2(T)}{T_c-T}~,~~~~T<T_c~~  
\end{eqnarray}
by substituton of $\varepsilon$ in Eq.~(\ref{si}) 
by $Tdp_i/dT - p_i(T)$ according to Eq.~(\ref{epsands}). 
Simple geometrical meaning of these inequalities is quite clear: they
are satisfied for any convex (from below) function
$p_i(T)$.  Any physical pressure function $p(T)$ should have positive
second derivative, $d^2p/dT^2>0$, and, therefore, is indeed a convex
function. To prove this last statement we use the relation
\begin{equation}\label{endir}
\frac{d^2p}{dT^2}~=~\frac{1}{T}\frac{d\varepsilon}{dT}~, 
\end{equation}
which follows from Eq.~(\ref{epsands}). Positive sign of 
$d\varepsilon/dT$ is a 
 consequence of the definition of energy in statistical
mechanics:  
\begin{equation}\label{energy} \varepsilon ~=~ \frac {\langle
E\rangle}{V}~=~ \frac{1}{V}~\frac{\sum_{n}E_n \exp(-E_n/T)} {\sum_{n}
\exp(-E_n/T)}~.  
\end{equation} 
~From Eq.~(\ref{energy}) one finds
\begin{equation}\label{positive}
\frac{d\varepsilon}{dT}~=~\frac{1}{V}~\frac{d\langle E\rangle}{dT}~=~
\frac{\langle E^2 \rangle - \langle E \rangle ^2}{VT^2}~=~ \frac{\langle
(E- \langle E\rangle)^2\rangle}{VT^2} ~>~0~.  
\end{equation}

\vspace{1cm}

{\bf Acknowledgements}

The results obtained by the NA35 and NA49 
Collaborations play a major role in the
interpretation of the whole set of data.
We would like to specially thank P.~Seyboth and R.~Stock 
spokesmen of these experiments.
We thank K.A.~Bugaev, D.~Ferenc, L.~Frankfurt, U.~Heinz, C.~Lourenco,
St.~Mr\'owczy\'nski, R.~Renfordt, H.~Str\"obele and B.~Svetitsky for
critical and vivid discussions and comments
to the manuscript.
M.I.G. is also grateful to the BMFT, DFG and GSI for the financial
support.

%\begin{figure}[p]
%\epsfig{file=ppe.eps,width=14cm}
%\caption{
%The $p/\epsilon$ ratio as a function of $F$.
%}
%\label{ppe}
%\end{figure}

\end{document}